\documentclass[12pt]{iopart}
\usepackage{iopams}  

\usepackage[latin1]{inputenc}
\usepackage{graphicx}
\usepackage{amssymb}
\usepackage{color}
\usepackage{float}
\expandafter\let\csname equation*\endcsname\relax 
\expandafter\let\csname endequation*\endcsname\relax
\usepackage{amsmath}
\usepackage{amsfonts}
\usepackage{dcolumn}
\usepackage{hyperref}
\usepackage{color}
\usepackage{enumerate}
\usepackage{subfigure}
\usepackage{epsfig}
\usepackage{epsf}
\usepackage{epstopdf}
\newcommand*{\everymodeprime}{\ensuremath{\prime}}
\usepackage{changepage}
\usepackage{blindtext}
\usepackage{enumitem}
\usepackage{tabularx}

\def\case#1/#2{\textstyle\frac{#1}{#2}}

\newcommand{\be}{\begin{eqnarray}}
\newcommand{\ee}{\end{eqnarray}}
\newcommand{\ben}{\begin{eqnarray}}
\newcommand{\een}{\end{eqnarray}}
\newtheorem{thm}{Theorem}

\newtheorem{defn}{Definition}

\begin{document}

\title{Global structure of static spherically symmetric solutions surrounded by quintessence}

\author{Miguel Cruz$^{1}$, Apratim Ganguly$^{2,3}$, Radouane Gannouji$^{4}$, Genly Leon$^{4}$, Emmanuel N. Saridakis$^{5,6}$}

\address{$^{1}$ Facultad de F\'\i sica, Universidad Veracruzana 91000, Xalapa, Veracruz, M\'exico}
\address{$^{2}$ Department of Mathematics, Rhodes University, 6140 Grahamstown, South Africa}
\address{$^{3}$ Astrophysics and Cosmology Research Unit, School of Mathematics, Statistics and Computer Sciences, University of KwaZulu-Natal, Private Bag X54001, Durban 4000, South Africa}
\address{$^{4}$ Instituto de F\'{\i}sica, Pontificia Universidad  Cat\'olica de Valpara\'{\i}so, Casilla 4950, Valpara\'{\i}so, Chile}
\address{$^{5}$ Physics Division, National Technical University of Athens, 15780 Zografou Campus, Athens, Greece}
\address{$^{6}$ CASPER, Physics Department, Baylor University, Waco, TX 76798-7310, USA}




\begin{abstract}
We investigate all static spherically symmetric solutions in the context of general relativity 
surrounded by a minimally-coupled quintessence field, using dynamical system analysis.
Applying the $1+1+2$ formalism and introducing suitable normalized variables 
involving the Gaussian curvature, we were able to reformulate the field equations as first order 
differential equations. In the case of a massless canonical scalar field we recovered all 
known black hole results, such as the Fisher solution, and we found that  apart from the 
Schwarzschild solution all other solutions are naked singularities. Additionally, we 
identified the symmetric phase space which corresponds to the white hole part of the 
solution and in the case of a phantom field, we were able to extract the 
conditions for the existence of wormholes and define all possible class of solutions such as Cold Black holes, singular spacetimes and wormholes like Ellis wormhole, for example. For an exponential potential, we found that the black hole solution which is asymptotically flat is unique and it is the 
Schwarzschild spacetime, while all other solutions are naked singularities. Furthermore, we 
found solutions connecting to a white hole through a maximum radius, and not a minimum radius (throat) such as wormhole solutions, therefore violating the flare-out condition. Finally, we have found a necessary and sufficient condition on the form of the potential to have an asymptotically AdS spacetime along with a necessary condition for the existence of asymptotically flat black holes.
\end{abstract}

\maketitle

\section{Introduction}

In general relativity singularities are formed through the collapse of massive objects, 
nevertheless under certain conditions these singularities are hidden behind a horizon, 
forming black holes. Is this simple picture universal? 

During the last decades, various situations have been studied in order to extend this 
analysis, from a canonical minimally coupled scalar field to higher spin fields. This work 
becomes more pertinent if we consider that any quantum theory of gravity introduces 
additional fields. Therefore, various extended versions of general relativity have been 
studied over the years.

In majority of such extended versions of gravity one obtains one extra scalar degree 
of freedom. For instance, in the case of $f(R)$ gravity, a conformal transformation allows 
to re-write the theory in the Einstein frame, i.e., to general relativity plus a scalar 
field. For brane models, the brane bending mode can describe the embedding of the brane in 
the bulk in a theory dubbed later on as Galileon \cite{Nicolis:2008in}. In massive gravity, 
an additional scalar field appears in the decoupling limit and it is responsible for the 
most interesting physics such as the vDVZ discontinuity or the Vainshtein mechanism 
\cite{deRham:2014zqa}. Additionally, scalar fields do also appear in various cases which 
include, e.g. compactified extra dimensions, such as the dilaton \cite{Gibbons:1987ps}.
Finally, and more generally, scalar fields appear in every quantum description of gravity, 
and these models are considered as phenomenological theories which can capture some of the 
details of a more fundamental theory, namely the effective field theories 
\cite{DeWitt:1967yk,Friedan:1985ge}. These scalar fields are known as dilaton, galileon, 
scalaron or just quintessence field. From the overstated, one understands the 
large success of studying scalar fields in various frameworks such as cosmology, black 
holes, screening mechanisms, spontaneous scalarization or superradiance. These models 
have also been extended to larger group of models like Horndeski \cite{Horndeski:1974wa}.

Hence, there has been a great effort in order to study the spherically symmetric and black
hole solutions in the framework of extensions of general relativity, and especially in the 
presence of an additional scalar field. According to a famous theorem by Bekenstein 
\cite{Bekenstein:1972ny}, a stationary canonical and minimally coupled scalar field which 
satisfies the condition $\phi V'(\phi)\ge 0$ is trivial, which means that the only regular black hole solution is the Schwarzschild spacetime. Following a slightly different 
proof, one may arrive at a different condition on the potential $V''(\phi)>0$ instead of $\phi V'(\phi)\ge 0$. This 
no-hair theorem suggests that stationary, asymptotically flat black holes for minimally 
coupled scalar tensor theory will not differ from black holes in general relativity. It 
has been later generalized to different models such as Brans-Dicke \cite{Hawking:1972qk}.

However, violating the assumptions of these theorems, although does not guarantee a non-trivial 
solution, may lead to new solutions with scalar hair. Such scalar field carries a conserved Noether charge
but not subject to a Gauss's law, unlike the electric charge, referred in the literature as hairy black holes 
\cite{Herdeiro:2015waa}. In some cases, this definition is extended to non-trivial
fields for which scalar charge is not an independent parameter. It has to be 
noticed that even if the no-hair theorems concerns the end point of a dynamical collapse, 
hairy black holes have been studied in a stationary framework.

Additionally, we need to keep in mind that working with a specific coordinate system 
might generate confusion, since two solutions that appear naively to be different may 
actually correspond to the same solution. In fact, as it is well known, the spherically 
symmetric solution for a massless scalar field was first discovered by Fisher 
\cite{Fisher:1948yn}, which we will call in the rest of the paper as the Fisher solution. 
However, it was unknown to the western world. Independently, Janis, Newman and Winicour 
rediscovered the solution \cite{Janis:1968zz} and later on another solution was found by 
Wyman \cite{Wyman:1981bd}. It is only in \cite{Virbhadra:1997ie} that it was noticed that 
these last two solutions are same and they correspond to the Fisher solution.

From the above discussion we deduce that it would be very useful to be able to obtain the 
global structure of a spacetime, without using a specific coordinate system. Hence, in 
the present work, we propose a new approach based on dynamical-system analysis in order 
to study the full spectrum of static spherically symmetric solutions in the case of general 
relativity with a minimally-coupled scalar field $\psi$, given by the action
\begin{eqnarray}
S=\int d^{4}x\sqrt{-g}\Bigg[\frac{R}{2}
- \frac{\varepsilon}{2} \partial_{\mu}\psi\partial^{\mu}\psi - V(\psi)\Bigg] 
\label{action},
\end{eqnarray}
where for generality we consider both the canonical ($\varepsilon=+1$) and phantom 
($\varepsilon=-1$) field cases. A similar analysis was previously introduced by some 
of us in \cite{Ganguly:2014qia}, which allowed to find various standard results in 
the context of general relativity without solving the field equations, such as the 
singularity-free nature of the Nariai solution. Nevertheless, in the present manuscript we 
extend this analysis in the presence of a scalar field.

The plan of the work is the following: In Sec. \ref{covarappr}, we review the basic 
equations of the $1+1+2$ covariant approach and introduce a set of dimensionless variables
along with the dynamical system describing a minimally coupled scalar field in Sec. \ref{dynamicalsystem}.
In Sec. \ref{masslesscase}, we investigate the massless, canonical or phantom scalar field in
detail and a complete analysis for an exponential potential is performed in Sec. \ref{Exponentialcase}. Finally,
in Sec. \ref{Generalcase}, we extract global information for a general potential before 
concluding in Sec. \ref{Conclusions}.
 
\section{1+1+2 covariant approach for spherically symmetric geometry}
\label{covarappr}

The $1+1+2$ decomposition \cite{Clarkson:2002jz,Clarkson:2007yp} follows the same 
strategy as the $1+3$ decomposition \cite{Ellis:1998ct,Ellis:1971pg}, where the spacetime
is decomposed into a timelike vector field and an orthogonal three-dimensional spacelike 
hypersurface. This surface is further decomposed into a spacelike vector field and
a 2-surface. Hence, all the informations are embedded in a set of kinematic and dynamical variables.
In cosmology, where one usually considers Friedmann-Robertson-Walker (FRW) spacetimes, the
3-dimensional hypersurface becomes homogeneous and isotropic. Therefore, only scalar fields 
remain after the decomposition, namely the expansion scalar $(\theta)$, the density
$(\rho)$ and the pressure $(P)$.

In this work we will consider the $1+1+2$ formalism for spherically symmetric spacetimes. 
The advantage of our formalism is that it allows for investigation without the need to 
consider an explicit coordinate choice for the metric, since it might lead to confusions \cite{Virbhadra:1997ie}. 
Another advantage of this formalism is that the full set of variables, after decomposition, 
will be scalars \cite{Clarkson:2002jz}, since there will be no privileged directions on 
the 2-surface for any spherically symmetric spacetime. The only non-zero variables for any locally rotationally symmetric
(LRS-II) spacetime are scalars in the $1+1+2$ approach.

\subsection{Formalism} 

Let us first perform the standard $1+3$ decomposition. We define a unit timelike vector 
$u^a$ $(u^a u_a=-1)$, which defines the projection tensor on the 3-space 
$h^a_b=g^a_b+u^au_b$. Therefore, we can 
define two derivatives, one  along the vector $u^{a}$ defined as
\begin{eqnarray}
\dot{T}^{a..b}{}_{c..d}{} = u^{e} \nabla_{e} {T}^{a..b}{}_{c..d}~, 
\end{eqnarray}
and a projected derivative defined as
\begin{eqnarray}
D_{e}T^{a..b}{}_{c..d}{} = h^a{}_f h^p{}_c...h^b{}_g h^q{}_d h^r{}_e \nabla_{r} 
{T}^{f..g}{}_{p..q}~.
\end{eqnarray}
Next, in order to perform the split of the 3-space, we introduce a unit 
spacelike vector $n^{a}$, such that
\begin{eqnarray}
n_{a} u^{a} = 0\;,\; \quad n_{a} n^{a} = 1,
\end{eqnarray}
along with a projection tensor on the 2-space (sheet) orthogonal to $n^a$ and $u^a$, 
\begin{eqnarray} 
N_{a}{}^{b} \equiv h_{a}{}^{b} - n_{a}n^{b} = g_{a}{}^{b} + u_{a}u^{b} 
- n_{a}n^{b}~,~~N^{a}{}_{a} = 2~.
\label{projT} 
\end{eqnarray} 
Thus, we can introduce two additional derivatives in the surface orthogonal 
to $u^a$, one along the vector $n^a$ 
\begin{eqnarray}
\hat{T}_{a..b}{}^{c..d} &\equiv  n^{f}D_{f}T_{a..b}{}^{c..d}~, 
\end{eqnarray}
and the corresponding projected derivative onto the sheet
\begin{eqnarray}
\delta_e T_{a..b}{}^{c..d} &\equiv 
N_{a}{}^{f}...N_{b}{}^gN_{i}{}^{c}..N_{j}{}^{d}N_e{}^kD_k T_{f..
g}{}^{i..j}\,.
\end{eqnarray}
%

\subsection{Variables} 

The set of variables is obtained by decomposing the various tensors along the timelike 
and spacelike directions. The energy-momentum tensor $T_{ab}$ can be decomposed
relative to $u^a$ as
\begin{eqnarray}
T_{ab}=\rho u_a u_b+p h_{ab}+q_b u_a+q_a u_b+\pi_{ab},
\end{eqnarray}
where $\rho$ is the energy density, $p$ the isotropic pressure, $q^a$ the energy flux and 
$\pi_{ab}$ the trace-free anisotropic pressure (anisotropic stress). After the 
decomposition of each tensor $(q_a,\pi_{ab})$ along the spacelike vector $n^a$, the only non-zero part of the heat flux and anisotropic pressure read 
as
\begin{eqnarray}
&&q_a = Q n_a,\nonumber\\
&&\pi_{ab}=\Pi (n_a n_b-\frac{1}{2}N_{ab}),
\end{eqnarray}
where $Q$ is the scalar part of the heat flux and $\Pi$ is the scalar part of the 
anisotropic stress.

Regarding the geometrical variables, the electric part of the Weyl tensor is decomposed as 
$E_{ab}=\mathcal{E}(n_a n_b-\frac{1}{2}N_{ab})$,  and since we focus on vorticity-free 
LRS-II spacetimes, the magnetic Weyl curvature becomes $H_{ab}=0$ \cite{Betschart:2004uu}.
The additional non-zero geometrical quantities are respectively the expansion
\begin{eqnarray}
 \Theta = \nabla_a u^a,
\end{eqnarray}
  the shear
\begin{eqnarray}
\Sigma = n^a n^b \nabla_{a} u_b,
\end{eqnarray}
the sheet expansion
\begin{eqnarray}
 \phi = \delta_a n^a,
\end{eqnarray}
and finally, the acceleration
\begin{eqnarray}
 \mathcal{A} = n^a \dot u_a.
\end{eqnarray}
%

\subsection{Equations} 

As we mentioned in the Introduction, in the present work we are interested in the static 
LRS class II spacetimes for the action (\ref{action}),  i.e., for general relativity with 
a minimally-coupled scalar field $\psi$. The advantage of our formalism is that it allows 
for investigation without the need to consider an explicit coordinate choice for the 
metric. Thus, for an unspecified metric $g_{ab}$, the energy-momentum tensor 
for the scalar field is given by
\begin{eqnarray}
\label{sf_1}
T_{a b}^{(\psi)}:= \varepsilon \nabla_a \psi \nabla_b \psi -\frac{1}{2}g_{a 
b}\left[\varepsilon \left(\nabla\psi\right)^2+2 
V(\psi)\right],
\end{eqnarray} 
where $V(\psi)$ is the scalar field potential. 

Considering only static spacetimes, all time derivatives are zero, which implies 
$\Theta=\Sigma=Q=0$ and $\dot\psi=0$ \cite{Ganguly:2014qia}. Hence, the 1+1+2 
decomposition of \eqref{sf_1} leads to
\begin{subequations}
\begin{eqnarray}
&\rho=\frac{1}{2}\varepsilon \hat\psi^2+V(\psi), \label{eq:rho}\\
&p= -\frac{1}{6}\varepsilon \hat\psi^2-V(\psi),\label{eq:P}\\
&\Pi =\frac{2}{3}\varepsilon \hat\psi^2,\label{eq:Pi}
\end{eqnarray}
\end{subequations}
where the hat $\hat{\ }$ marks the derivative along the spacelike vector 
(e.g. $\hat\psi=n^\mu D_\mu \psi$). From which we can easily derive the propagation and constraint
equations \cite{Clarkson:2002jz}
\begin{subequations}
\begin{eqnarray} 
&\hat\phi = -\frac{1}{2}\phi^2-\frac{2}{3}\rho-\frac{\Pi}{2}-{\cal E}~,
\label{equation13}\\
&\hat{\cal E} -\frac{\hat\rho}{3} + \frac{\hat\Pi}{2} =- \frac{3}{2}\phi\left({\cal 
E}+\frac{1}{2}\Pi \right),
\label{equation14}\\
&\hat{\cal A} = -\left({\cal A}+\phi\right){\cal A} +\frac{1}{2}\left(\rho +3p \right)~, 
\label{equation15}\\
&\hat p+\hat\Pi = -\left(\frac{3}{2}\phi+{\cal A}\right)\Pi -\left(\rho+p\right){\cal A}~,
\label{equation16}
\end{eqnarray}
\end{subequations}
and
\begin{subequations}
\begin{eqnarray}  
&0 = - {\cal A}\phi +\frac{1}{3} \left(\rho +3p  \right) -{\cal E} +\frac{1}{2}\Pi~,
\label{equation17}\\ 
& K = \frac{\rho}{3}-{\cal E}-\frac{\Pi}{2}+\frac{\phi^{2}}{4},\label{equation18}
\end{eqnarray} 
\end{subequations}  
where we have defined the Gaussian curvature via the Ricci tensor on the sheet as
$^2R_{ab}=K N_{ab}$ \cite{Betschart:2004uu}. Taking the spatial derivative of \eqref{equation18}, 
and using the previous equations, we get
\begin{eqnarray}
\hat K &= -\phi K ~. \label{equation19}
\end{eqnarray}
Note that the constraint equations, which are equivalent to the 
hamiltonian and momentum constraints, include no derivatives.

Introducing a new variable, $\Psi=\hat\psi$, we can rewrite the system \eqref{equation13}-\eqref{equation16} with the help of \eqref{eq:rho}-\eqref{eq:Pi} and \eqref{equation19}, in the following form
\begin{subequations}
\begin{eqnarray}
\hat\psi &=\Psi,\label{equation20}\\
\hat\phi &= -\frac{1}{2}\phi^2-\frac{2}{3}\left[\varepsilon \Psi^2+V(\psi) \right]-{\cal E}~,
\label{equation21}\\
\hat{\cal E}  &= \frac{\varepsilon }{3}\Psi^2\left({\cal A}-\frac{1}{2}\phi\right)- 
\frac{3}{2}\phi 
{\cal E}~,
\label{equation22}\\
\hat{\cal A} &= -\left({\cal A}+\phi\right){\cal A} -V(\psi) 
\label{equation23}\\
\varepsilon \hat\Psi &= -\varepsilon \left({\cal A}+\phi\right)\Psi +V'(\psi),
\label{equation24}\\
\hat K &= -\phi K ~, \label{equation25}
\end{eqnarray}
\end{subequations}
subject to the constraints (\ref{equation17},\ref{equation18})
\begin{subequations}
\begin{eqnarray}  
&{\cal E} = -{\cal A} \phi -\frac{2 V(\psi )}{3}+\varepsilon \frac{\Psi ^2}{3},
\label{equation26}\\ 
& K = {\cal A} \phi +V(\psi )-\varepsilon \frac{\Psi ^2}{2}+\frac{\phi 
^2}{4}.
\label{equation27}
\end{eqnarray}   
\end{subequations}
\newpage
\section{The dynamical system}
\label{dynamicalsystem}

Let us now follow the standard procedure and re-write the system of equations as an 
autonomous dynamical system. Similar to the cosmological case, where one uses the  
Friedmann equations in order to define the auxiliary dimensionless variables   
($\Omega_m,\Omega_r,\cdots$), in the present case we will use Eq. (\ref{equation27}) 
in order to introduce the suitable dimensionless variables
\begin{eqnarray}
x_1=-\frac{{\cal E}}{K},\;\; x_2= \frac{\phi}{2 \sqrt{K}}, \;\; x_3=\frac{{\cal 
A}}{\sqrt{K}},\;\; y_1=\frac{\Psi}{\sqrt{2 K}},\;\; y_2=\frac{V(\psi)}{3 K}.
\label{massive_variables}
\end{eqnarray}
Therefore, the constraints equations  (\ref{equation26}),(\ref{equation27}) become
\begin{subequations}
\begin{eqnarray}
\label{Gauss_2}
& x_2^2+2 x_2 x_3-\varepsilon y_1^2+3 y_2=1,
\\
\label{E2}
& 3 x_1-6 x_2 x_3+2 \varepsilon y_1^2-6 y_2=0.
\end{eqnarray}
\end{subequations}
Furthermore, we define the additional variables
\begin{eqnarray}
\lambda=-\frac{V_{,\psi}}{V},\; \Gamma=\frac{V V_{,\psi\psi}}{V_{,\psi}^2},
\end{eqnarray}
and thus for a given potential the scalar field can be expressed as a function of 
$\lambda$ ($\psi=\psi(\lambda)$), or equivalently of $\Gamma=\Gamma(\lambda)$.
In summary, the propagation equations for the variables \eqref{massive_variables}
and $\lambda$ are given by
\begin{subequations}
\label{syst_24}
\begin{eqnarray}
&x_1'=\frac{2}{3} \varepsilon y_1^2 (x_2-x_3)-x_1 x_2,\\
&x_2'=\frac{1}{6} \left(3 x_1-4 \varepsilon y_1^2-6 y_2\right),\\
&x_3'=-x_3
   (x_2+x_3)-3 y_2,\\
&y_1'=-y_1 (x_2+x_3)-\varepsilon \frac{3 \lambda  y_2}{\sqrt{2}},\\
&y_2'=y_2 \left(2 x_2-\sqrt{2} \lambda
    y_1\right),\label{syst_24e}\\
&\lambda'=-\sqrt{2} (\Gamma -1) \lambda ^2 y_1,
\end{eqnarray}
\end{subequations}
where primes denote the normalized spatial derivative $f'=\frac{\hat f}{\sqrt{K}}$ 
(where as we have said the hat $\hat{\ }$ marks the derivative along the spacelike vector 
(e.g. $\hat\psi=n^\mu D_\mu \psi$)). In the case where one chooses a particular 
coordinate system, the above derivatives become derivatives with respect to the radial 
coordinate, however the present formalism helps to handle the system in a coordinate-independent manner. 

From \eqref{syst_24e} it follows that the sign of $y_2$ (i.e., the sign of $V(\psi)$) is invariant for the flow. This implies that our set of variables is not suitable for models where the potential changes sign. For this set of variables, the potential should be always positive or always negative because of the definition of $\lambda=-V_{,\psi}/V$ which diverges when $V=0$.
Furthermore, the sign of $\lambda$ and $\Gamma$ remain unaffected under the change $V(\psi)\rightarrow -V(\psi)$ (or in other words, they remain unaffected under the change $y_2\rightarrow -y_2$). In the following we will investigate the massless case $y_2=0$ and then, we will study the case $y_2\neq 0$, with special mention of some specific solutions for negative potential ($y_2<0$), although, the main attention will be restricted to positive potentials ($y_2>0$).

Since the constraints \eqref{Gauss_2} and \eqref{E2} are conserved, we can use 
them in order to eliminate two variables, for instance $x_1$ and $y_2$, which leads to 
the reduced dynamical system
\begin{subequations}
\label{general_Potential}
\begin{eqnarray}
&x_2'=x_2 x_3-\varepsilon y_1^2,\\
&x_3'=x_2^2+x_2 x_3-x_3^2-\varepsilon y_1^2-1,\\
&y_1'=\varepsilon \frac{\lambda \left(x_2^2+2 x_2 x_3-\varepsilon y_1^2-1\right)}{\sqrt{2}}-y_1 
(x_2+x_3),
\\
&\lambda'=-\sqrt{2} (\Gamma -1) \lambda ^2 y_1.
\end{eqnarray}
\end{subequations}
We note that for a positive potential $(y_2>0)$, we obtain
the condition $x_2 (x_2+2 x_3)-\varepsilon y_1^2\leq 1$ from (\ref{Gauss_2}). Therefore, for non-negative 
potentials the above system defines a flow on the unbounded phase space
\begin{eqnarray}\label{phase_space_gen}
\left\{(x_2,x_3,y_1,\lambda): x_2 (x_2+2 x_3)-\varepsilon y_1^2\leq 1, 
\lambda\in\mathbb{R}\right\}.
\end{eqnarray}
On the other hand, for negative potentials, we have to consider the phase space 
\begin{eqnarray}
\label{phase_space_gen_negative}
\left\{(x_2,x_3,y_1,\lambda): x_2 (x_2+2 x_3)-\varepsilon y_1^2\geq 1, 
\lambda\in\mathbb{R}\right\}.
\end{eqnarray}
Defining $f(\lambda)=(\Gamma(\lambda) -1) \lambda ^2$, where we have assumed that $\Gamma$ can be expressed as a function of $\lambda$ as displayed in Table \ref{fsform} \cite{Escobar:2013js} (see references therein), we can examine different classes of potentials. In the following sections, we will investigate various cases of specific potentials 
separately.

\Table{\label{fsform} The function  $f(\lambda)$ for the most common quintessence potentials \cite{Escobar:2013js} (see references therein).}
\br
\ns
Potential & $f(\lambda)$ \\
\mr 
$V(\psi)=V_0 \psi^N$ & $-\frac{\lambda^2}{N}$ \\
$V(\psi)=V_{0}e^{-k\psi}+V_1, V_1>0$ & $-\lambda(\lambda-k)$ \\
$V(\psi)=V_{0}\left[e^{\alpha\psi}+e^{\beta\psi}\right]$ &	$-(\lambda+\alpha)(\lambda+\beta)$ \\
$V(\psi)=V_{0}\left[\cosh\left( \xi \psi \right)-1\right]$ &  $-\frac{1}{2}(\lambda^2-\xi^2)$ \\
$V(\psi)=V_{0}\sinh^{-\alpha}(\beta\psi), \alpha>0$ & $\frac{\lambda^2}{\alpha}-\alpha\beta^2$ \\
\br
\end{tabular}

\end{indented}
\end{table}

\section{Massless scalar field}
\label{masslesscase}

\subsection{Canonical field}

In this section, we will study a standard solution, the Fisher solution \cite{Fisher:1948yn},
which corresponds to a massless canonical scalar field (i.e., with $\varepsilon =1$ and $V(\psi)=0$).
We will see that, using dynamical-system approach, we can recover all standard results of
this solution without explicitly solving the equations.

For our auxiliary variables, a null potential $V=0$ implies $y_2=\lambda=0$. Moreover,
from  Eq. (\ref{Gauss_2}), we obtain an additional constraint for this particular case, 
$y_1^2=x_2^2+2x_2x_3-1$, which allows us to reduce the dynamical system to 2D. 
The phase space is given by $(x_2,x_3)$ with a constraint defined as $y_1^2=x_2^2+2x_2x_3-1>0$, 
which implies a real scalar field. In fact, $y_1^2=\hat{\psi}^2/2K$
is positive for any real valued scalar field since $K>0$ ($K=1/r^2$ in 
Schwarzschild coordinates).

In summary, we obtain the reduced dynamical system
\begin{subequations}
\label{syst_1}
\begin{eqnarray}
&x_2'=1-x_2(x_2+x_3),\\
&x_3'=-x_3(x_2+x_3),
\end{eqnarray}
\end{subequations}
defined in the phase space
\begin{eqnarray} 
\left\{(x_2,x_3): x_2^2+2x_2 x_3\geq 1\right\}.
\end{eqnarray}
The system \eqref{syst_1} admits two fixed points in the finite region, whose stability 
can be obtained by following the usual linearization procedure and examining the 
eigenvalues of the involved perturbation matrix. In particular, these are:
\begin{enumerate}
\item[$P_M$:] $(x_2=1, x_3=0)$.  The corresponding eigenvalues are $-2,-1$, and thus 
 this point is a local sink, i.e., stable point.
\item[$\bar P_M$:] $(x_2=-1, x_3=0)$. The corresponding eigenvalues are $2,1$, and thus this 
point is a local source, i.e., unstable point. 
\end{enumerate}
We can easily reconstruct the metric at these points. Specifically, following 
\cite{Ganguly:2014qia} we have
$B=x_2^2$ and $d\ln A/d\ln r=2x_3/x_2$, where $(A,B)$ are the gravitational potentials 
defined as 
\begin{eqnarray}
{\rm ds}^2=-A(r){\rm dt}^2+\frac{{\rm dr}^2}{B(r)}+r^2\Bigl({\rm d\theta}^2+\sin^2\theta{\rm 
d\phi}^2\Bigr).
\end{eqnarray}
We, therefore, see that the two critical points correspond to Minkowski spacetime, since, 
$B=x_2^2=1$ and $d\ln A/d\ln r=2x_3/x_2=0$, which implies $A=const.$ and 
can be set to $A=1$ by a time redefinition.

Since the system is defined on an unbounded phase space, there may exist non-trivial 
behavior at the region where the variables diverge. For this reason, we need to introduce 
Poincar\'e variables to study the behavior of the system at infinity, such as,
\begin{subequations}
\begin{eqnarray}
\label{Poincarea}
X_2&=\frac{x_2}{\sqrt{1+x_2^2+x_3^2}},\\
X_3&=\frac{x_3}{\sqrt{1+x_2^2+x_3^2}},
\label{Poincareb}
\end{eqnarray} 
\end{subequations}
with inverse transformation:
\begin{subequations}
\begin{eqnarray}
x_2&=\frac{X_2}{\sqrt{1-X_2^2-X_3^2}},\\
x_3&=\frac{X_3}{\sqrt{1-X_2^2-X_3^2}}.
\end{eqnarray} 
\end{subequations}
The infinite boundary $x_2^2+x_3^2\rightarrow +\infty$ corresponds to the unitary circle 
$X_2^2+X_3^2=1$. The propagation equations read as 
\begin{subequations}
\label{Poincare_1}
\begin{eqnarray}
&\tilde X_2= \left[1-X_2  \left(2 X_2 +X_3 \right)\right]\left(1-X_2 ^2-X_3 ^2\right),\\
&\tilde X_3=-X_3 \left( 2 X_2 +X_3 \right) \left(1-X_2 ^2-X_3 ^2\right),
\end{eqnarray}
\end{subequations}
defined on the phase space 
\begin{eqnarray}
\left\{(X_2,X_3): 2 X_2^2+2 X_2 X_3+X_3^2\geq 1, X_2^2+X_3^2\leq 1\right\},
\end{eqnarray}
where we have rescaled the radial variable as 
\begin{eqnarray}
\tilde f\rightarrow \sqrt{1-X_2^2-X_3^2}f'.
\end{eqnarray}
In Table \ref{Table1}, we present the critical points at infinity 
for the Poincar\'e (global) system \eqref{Poincare_1}. 

\Table{\label{Table1} Critical points at infinity for the Poincar\'e (global) system \eqref{Poincare_1}, for the case of a massless canonical scalar field.}
\br
\ns
Point & $X_2$ & $X_3$ & Stability & Nature\\
\mr 
$P_H$  & $0$& $1$ & unstable & Horizon \\
$\bar P_H$  & $0$& $-1$ & stable & Horizon\\
$P_S$  & $\frac{2}{\sqrt{5}}$ & $-\frac{1}{\sqrt{5}}$& unstable & Singularity\\
$\bar P_S$ & $-\frac{2}{\sqrt{5}}$ & $\frac{1}{\sqrt{5}}$ & stable & Singularity\\
\br
\end{tabular}

\end{indented}
\end{table}

\begin{figure}[htb!]
\centering
\includegraphics[width=0.45\textwidth]{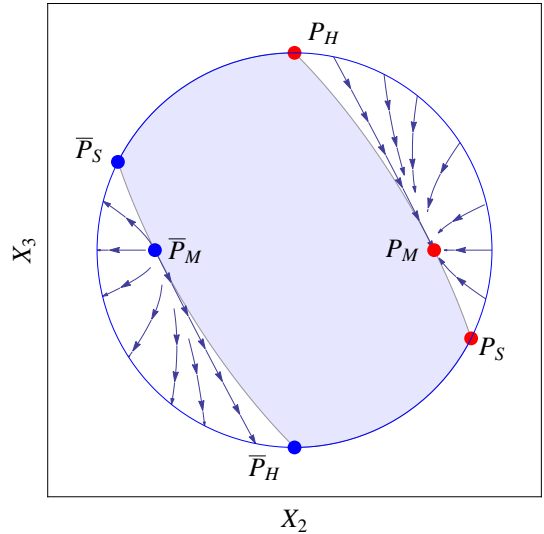}
\caption{\label{fig:Fig1} Global phase space for the system \eqref{Poincare_1},  
corresponding to the case of a massless canonical scalar field. The blue region is 
forbidden since it leads to the violation of the reality condition $y_1^2\geq 0$.}
\end{figure}
Additionally, in Fig. \ref{fig:Fig1}, we depict the corresponding (global) phase 
space behavior for the system \eqref{Poincare_1}, i.e., the global behavior of the 
spherically symmetric solutions in the case of a massless canonical scalar field. This 
phase space is 2D, therefore any trajectory (or any solution) demands two initial 
conditions, which can be related to two integration constants of the system, which we can 
call the mass 
and the scalar charge. The light blue region is forbidden since it leads to the violation 
of the reality condition $y_1^2\geq 0$. The critical points $(P_H,\bar P_H)$ correspond to 
horizons, since their $X_2$ coordinate is $X_2=0$ (see
\cite{Ganguly:2014qia} for the definition and \ref{PHmassless} for the stability). Finally, points $P_S$ and $\bar P_S$ 
correspond to singularity, since from the coordinates $d\ln A/d\ln 
r=2x_3/x_2=2X_3/X_2=-1$, which reduces to $A\propto 1/r$. The same behavior appears 
for B, since for the critical point, $B=x_2^2=\infty$, but after the linearization of the 
system following \cite{Ganguly:2014qia}, it appears that $B$ diverges as $1/r$ when $r\rightarrow 0$ (see \ref{PSmassless} for complete analysis of the stability). 

Since the line at infinity is a critical line, there is only one trajectory 
coming from each point of this line. In fact, by performing a linearization of the 
equations near the point $P_H$, we find $X_2+X_3\simeq 1$. This relation doesn't have a 
free parameter and hence it is unique, which implies that there is only one trajectory in 
the phase space from $P_H$. It is then easy to conclude that we have only one trajectory 
and, therefore, only one solution connecting the horizon $P_H$ to Minkowski point $P_M$
asymptotically. This is illustrated in Fig. \ref{fig:Fig1}, where we see only one 
trajectory connecting $P_H$ and $P_M$. This is the Schwarzschild solution, as it was
shown in \cite{Ganguly:2014qia}. Any other trajectory starts from a singularity and not 
from the horizon, and therefore it describes a naked singularity. In summary, the only 
solution that describes a black hole is the Schwarzschild solution. This is one of the 
most important properties of the Fisher solution: the unique solution describing a black 
hole is the trivial case where the scalar field is constant and the metric is 
Schwarzschild. Any other situation corresponds to a naked singularity. 

Let us now estimate how the metric diverges near the critical surface in the Poincar\'e 
coordinates. The solutions at infinity correspond to $x_2=\infty$ and leads to a 
divergence of the gravitational potential $B$. More precisely, we have 
$B=x_2^2=X_2^2/(1-X_2^2-X_3^2)$, hence, all points at infinity which correspond in 
Poincar\'e coordinates to the points on the circle $X_2^2+X_3^2=1$, have a divergent 
potential. To estimate how this divergence behaves, we can linearize the 
equations (\ref{Poincare_1}) as $X_2=\bar X_2-\varepsilon_2$ and $X_3=\bar X_3-\varepsilon_3$ 
where $\bar 
X_2^2+\bar 
X_3^2=1$, which gives for the region $X_2>0$
\begin{eqnarray}
\widetilde{\left(\bar X_2 \varepsilon_2+\bar X_3 \varepsilon_3\right) }=2\Bigl(\bar X_3+\sqrt{1-\bar 
X_3^2}\Bigr) \Bigl(\bar X_2 \varepsilon_2+\bar X_3 \varepsilon_3\Bigr).
\end{eqnarray}
And since $\tilde X=X_2 d X/d\ln r$ (see \cite{Ganguly:2014qia}), we have
\begin{eqnarray}
\frac{d\ln \Bigl(\bar X_2 \varepsilon_2+\bar X_3 \varepsilon_3\Bigr)}{d\ln 
r}=2\Bigl(1+\frac{\bar X_3}{\sqrt{1-\bar X_3^2}}\Bigr),
\end{eqnarray}
from which we can conclude that
\begin{eqnarray}
B\simeq r^{-2 \Bigl(1+\frac{\bar{X}_3}{\sqrt{1-\bar X_3^2}}\Bigr)}\quad \text{for}\quad
-\frac{1}{\sqrt{5}}<\bar X_
3 < 1,
\end{eqnarray}
or equivalently $B\simeq r^\alpha$ where $\alpha\leq -1$. Finally, we have $d\ln A/d\ln 
r=2x_3/x_
2=2X_3/X_2=2\bar{X}_3/\sqrt{1-\bar{X}_3^2}$, which implies $A\simeq 
r^{2\bar{X}_3/\sqrt{1-\bar{X}_
3^2}}$, or equivalently $A\simeq r^\beta$ with $\beta\geq -1$.

As we have noticed (see Fig. \ref{fig:Fig1}), there is a forbidden region corresponding 
to the violation of the reality condition by the scalar field. Since the Fisher solution can 
be written in the following form
\begin{eqnarray}
ds^2 &=-F^S dt^2+\frac{dr^2}{F^S}+r^2F^{1-S}d\Omega^2\nonumber\\
\psi &=\sqrt{\frac{1-S^2}{2}} \log F,\qquad F=1-\frac{r_S}{r},
\label{MFisher}
\end{eqnarray}
then the forbidden region corresponds to $S^2>1$. 

From the above discussion we see that $x_3=0$ is an invariant submanifold, and therefore 
it is interesting to study this case in particular. The equations reduce to
\begin{eqnarray}
x_2'=1-x_2^2\label{eq:Fisher}.
\end{eqnarray}
Knowing that the derivative $\everymodeprime$ is related to the radial coordinate in 
Schwarzschild 
coordinates as
\begin{eqnarray}
x' &\equiv \frac{dx}{d\xi} \quad \text{(where $\xi$ is an affine parameter)}\\
&=x_2 \frac{dx}{d\ln r} \quad \text{(see \cite{Ganguly:2014qia})},
\label{aux22}
\end{eqnarray}
it implies $x_2 d\xi=d\ln r$. Hence, from (\ref{eq:Fisher}) we have 
$x_2=(e^{2\xi}+\alpha)/(e^{2\xi}-
\alpha)$, where $\alpha$ is an integration constant, which  using (\ref{aux22}) implies 
that $r=e^{\xi}-\alpha e^{-\xi}$ and therefore 
\begin{eqnarray}
B=x_2^2=1+\frac{4\alpha}{r^2}.
\end{eqnarray}
Furthermore, since $x_3=0$ we deduce that $A=const.$, which can be set to  $A=1$ 
by a time redefinition, and this leads to the metric
\begin{eqnarray}
\label{Ellis}
ds^2=-dt^2+\frac{dr^2}{1+4\alpha/r^2}+r^2d\Omega^2.
\end{eqnarray}
Finally, since $y_1^2=4\alpha/r^2>0$, we have $\alpha>0$ and therefore this solution 
describes a naked singularity. This solution can be easily written in the form of the 
Fisher solution (\ref{MFisher}) when $S=0$. As shown previously, all non trivial 
solutions describe a naked singularity.

In conclusion, we see that the phase space of Fig. \ref{fig:Fig1} represents completely 
the spectrum of solutions in the case of a massless scalar field. We have obtained the 
solution connecting the horizon to an asymptotic Minkowski spacetime $(P_H\rightarrow 
P_M)$, which corresponds to the Schwarzschild black hole solution. Furthermore, we have 
the solution connecting a singularity to an asymptotically flat solution 
$(P_S\rightarrow P_M)$, corresponding to the Schwarzschild solution with negative mass 
(naked singularity, see \cite{Ganguly:2014qia}). All other solutions connect a singularity to the Minkowski 
spacetime. Therefore, all solutions are asymptotically flat, nevertheless they all 
include naked singularities, apart from the Schwarzschild solution.

Finally, we can easily see from Fig. \ref{fig:Fig1} that the phase space is divided 
in two parts. In the above, we have analyzed only the region $x_2>0$, since we can  easily 
see that the phase space region $x_2<0$ is similar but with opposite behavior. In 
particular, a critical point which is stable in one region becomes 
unstable in the other region and vice versa. As it was shown in \cite{Ganguly:2014qia}, 
the region $x_2<0$ corresponds to the white hole solution \cite{Haggard:2014rza}. Lastly, 
note that in this particular case of a massless scalar field, the two solutions are 
disconnected. 

\subsection{Phantom field}

\begin{figure}[htb!]
	\centering
		\includegraphics[width=0.45\textwidth]{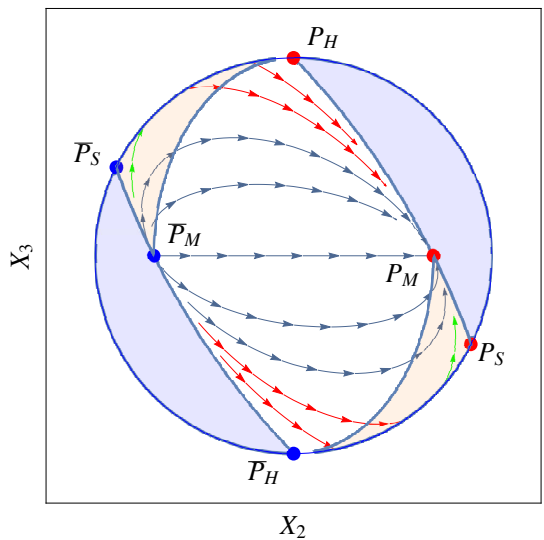}
	\caption{\label{fig:Fig2} Global phase space for a massless ghost scalar field. The blue 
region represents the forbidden part of the phase space. The orange region is defined by a negative 
Misner-Sharp mass.}
\end{figure}

For completeness, let us now investigate the case of a massless phantom scalar field, 
which has a negative kinetic term, i.e., taking $\varepsilon =-1$ in action (\ref{action}).
In this case the equations will be exactly the same, however the forbidden region will 
be the opposite, and the whole phase space will be the dual space of the canonical case, 
as can be seen in Fig. \ref{fig:Fig2}. The simplest of these solutions is known in 
the literature as the anti-Fisher solution, firstly found by Bergmann and Leipnik 
\cite{Bergmann:1957zza}. All solutions which go through $x_2=0$, where $x_2'=y_1^2>0$, 
satisfy the flare-out condition (see \ref{appendix:flare}), which implies a 
minimal size of the radius, namely a throat at $x_2=0$.

As we observe in Fig. \ref{fig:Fig2}, the phase space has a rich structure, and we have 
three types of solutions:
\begin{itemize}
\item The red curve represents a solution for which one asymptotic is Minkowski, while 
the other starts from the region $x_2<0$ for $x_3>0$. We see that asymptotically 
(near $P_M$) the Misner-Sharp mass is positive (see \ref{appendix:mass}). These 
solutions are known in the literature as Cold Black Holes \cite{Tremblay:2016ijg}.

\item The green curve represents solutions which are asymptotically Minkowski but start 
from the region $x_2>0$ for $x_3<0$. We can see that these solutions have a negative 
Misner-Sharp mass and are very similar to the Fisher solution. They exhibit a 
naked singularity.

\item The blue curves represent solutions where both asymptotics are Minkowski, with a 
minimum radius when $x_2=0$ (a throat). These are wormhole solutions
\cite{Visser:1995cc}. The particular case where $x_3=0$ gives the solution 
(\ref{Ellis}), however now with $\alpha<0$, which is known as the Ellis 
wormhole.  
\end{itemize}
%

\section{Exponential potential: $V=V_0 e^{-\lambda \psi}$}
\label{Exponentialcase}

In this section we will investigate the case of an exponential potential $V=V_0 e^{-\lambda \psi}$, where we have chosen $V_0\geq 0$, focusing on the 
standard case where the scalar field is non phantom, i.e where $\varepsilon=+1$. In the 
exponential potential case we have that $\lambda=-V_{,\psi}/V=const.$ and hence
$\Gamma=1$. Thus, the last 
equation of (\ref{general_Potential}) is trivially satisfied, and hence we can study the 
reduced 3D system for $(x_2,x_3,y_1)$ where $\lambda$ is a constant.

The fixed points at the finite region of the phase space are:
\begin{enumerate}
\item[$P_M$:] $(x_2=1, x_3=0,y_1=0)$. The corresponding eigenvalues are $2,-1,-1$, and 
therefore this point is saddle.
\item[$\bar P_M$:] $(x_2=-1, x_3=0,y_1=0)$. The corresponding eigenvalues are $-2,1,1$, 
and thus this point is saddle. 
\end{enumerate}
These two critical points represent Minkowski spacetime. However, and in significant 
contrast with the massless scalar case, these points are now saddle. Therefore it is not trivial to have an asymptotically flat spacetime (which should be an attractor). Only a reduced phase space will give rise to an asymptotically flat solution. And we will see that it corresponds to Schwarzschild spacetime.

\begin{figure}[htb!]
\centering
\includegraphics[width=.8\textwidth]{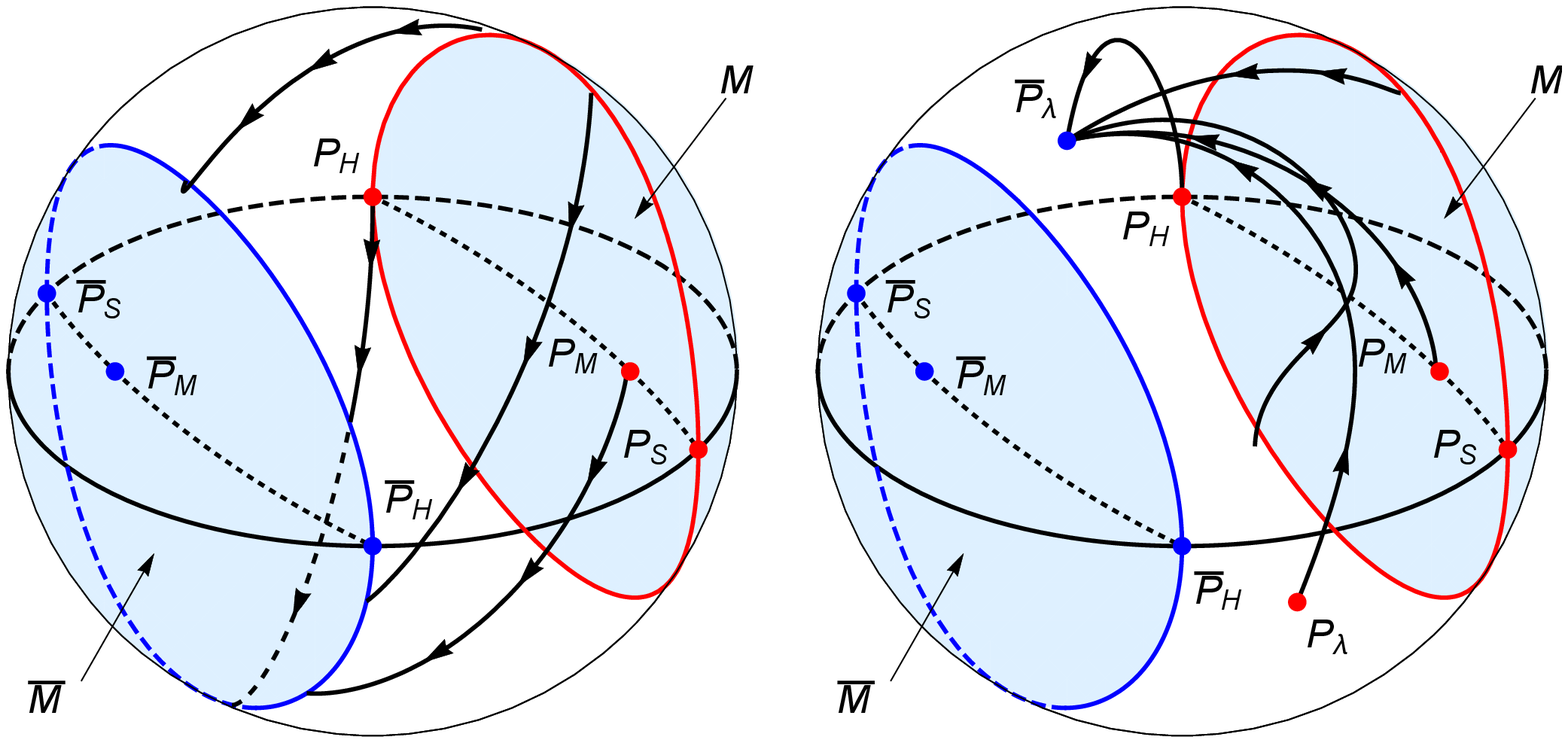}
\includegraphics[width=.15\textwidth]{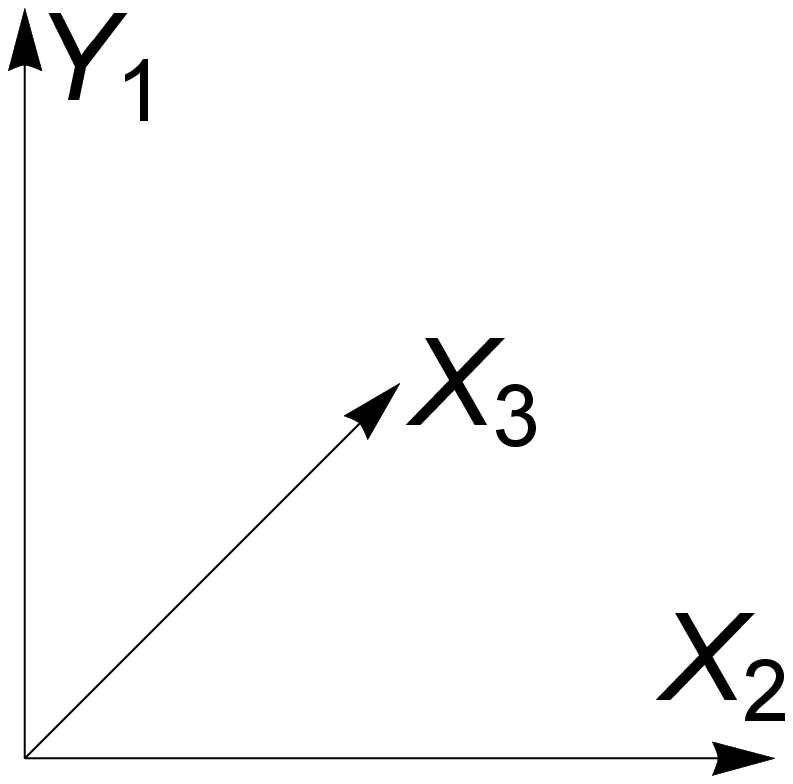}
\caption{\label{Fig:exp3D} Phase space of the exponential potential where the blue part represents the 
forbidden region (phantom scalar field). $M$ and $\bar M$ are the boundaries of the phase space and 
represent two invariant submanifold where most of the critical points are localized. On the left figure, the phase space is represented with some orbits for $\lambda=1$ for which all critical points are localized on $M$ and $\bar M$, while the right figure represents the orbits for $\lambda=-4$ and therefore $P_\lambda$ and $\bar P_\lambda$ exist.}
\end{figure}

Similarly to the previous section, we introduce the Poincar\'e variables
\begin{subequations} 
\label{Poincare_exp}
\begin{eqnarray}
\label{Poincareaa}
X_2&=\frac{x_2}{\sqrt{1+x_2^2+x_3^2+y_1^2}},\\
X_3&=\frac{x_3}{\sqrt{1+x_2^2+x_3^2+y_1^2}}, \label{Poincarebb}\\
Y_1&=\frac{y_1}{\sqrt{1+x_2^2+x_3^2+y_1^2}},
\label{Poincarecc}
\end{eqnarray} 
\end{subequations}
in order to study the phase space behavior at infinity. The infinity boundary $x_2^2+x_3^2+y_1^2\rightarrow +\infty$ corresponds to the unitary circle 
$X_2^2+X_3^2+Y_1^2=1$.\\
Their evolution equations become
\begin{subequations}
\label{syst44}
\begin{align}
\tilde X_2 &=-\frac{\lambda}{\sqrt{2}}  X_2 Y_1 \left[2 X_2^2+2 X_2 X_3+X_3^2-1\right] +Y_1^2 \left[X_2 (2 X_2+X_3)-1\right] \nonumber\\
&   -X_2 X_3 \left[X_2 (3 X_2+X_3)-2\right],\\
\tilde X_3&=-\frac{\lambda}{\sqrt{2}}  X_3 Y_1 \left[2 X_2^2+2 X_2  X_3+X_3^2-1\right]-3 X_2^2 X_3^2+2 X_2^2-X_2 X_3^3 \nonumber\\
 &  +X_3 Y_1^2 (2X_2+X_3)+X_2 X_3+X_3^2-1,\\
\tilde Y_1&=\frac{\lambda}{\sqrt{2}} \left[1-Y_1^2\right] \left[2 X_2^2+2 X_2   X_3+X_3^2-1\right]-X_2 Y_1 \left[3 X_2 X_3+X_3^2+1\right]\nonumber\\
   &+Y_1^3 \left[2X_2+X_3\right].
\end{align}
\end{subequations}
defined on the phase space 
\begin{eqnarray}
\left\{(X_2,X_3,Y_1):2 X_2^2+2 X_2 X_3+X_3^2\leq 1,  X_2^2+X_3^2+Y_1^2\leq 1\right\},
\end{eqnarray}
where we have rescaled the radial variable through 
\begin{eqnarray}
\tilde{X}\rightarrow \sqrt{1-X_2^2-X_3^2-Y_1^2}~X'.
\end{eqnarray}
Defining $\mathcal{I}(X_2,X_3,Y_1)=2 X_2^2+2 X_2 X_3+X_3^2-1$, it follows
\begin{eqnarray}
\tilde{\mathcal{I}}=\mathcal{I}\Bigl[-\sqrt{2} \lambda  Y_1 \left(2 X_2^2+2 X_2 
X_3+X_3^2\right)  -2 X_2 \left(3 X_2 X_3+X_3^2-1\right)\nonumber\\
+2 Y_1^2 (2 X_2+X_3)+2 X_3\Bigr].
\end{eqnarray}
Thus, $2 X_2^2+2 X_2 X_3+X_3^2= 1$ ($\mathcal{I}=0$) defines an invariant submanifold, 
which implies  
that any trajectory along this surface remains on the surface. These surfaces are denoted 
$M$ and $\bar M$ in the Fig. \ref{Fig:exp3D}.
%

The analysis of the above system shows that we have the same critical points than the 
massless case representing the horizon and the singularity (see \ref{PHexp} and \ref{PSexp} for complete analysis). Furthermore, 
we have an additional critical curve (i.e. curve of critical points), which is defined by 
the boundary of the phase space $2 X_2^2+2 X_2 X_3+X_3^2= 1$ and $X_2^2+X_
3^2+Y_1^2=1$ (marked by the red curve in Fig. \ref{Fig:exp3D}  for $x_2>0$ and the blue curve for $x_2<0$). We therefore 
see that most of the points, at infinity and at finite distance, are localized on the same 
critical surface $2 X_2^2+2 X_2 X_3+X_3^2= 1$.  

Hence, we can analyze two cases:
\begin{itemize}
\item The trajectory is on the critical surface ($M$ or $\bar M$). This case reduces 
to a 2D system studied below.

\item The trajectory flows from one surface to the other ($M$ to $\bar M$).
\end{itemize}

\begin{figure}[htb!]
\centering
\includegraphics[scale=.7]{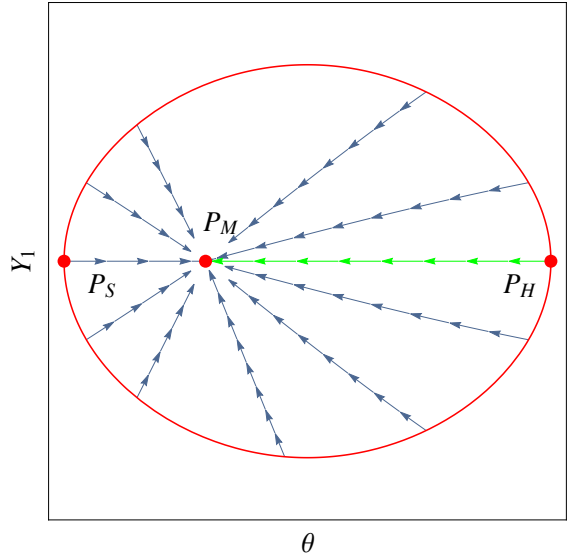}
\caption{\label{Figexp} Phase space of of the system (\ref{eqexp1})-(\ref{eqexp2}) representing the 
dynamics on the invariant surface $M$, for the case of exponential potential.}
\end{figure}

For the first case, and having in mind that $\bar M$ is just the inverse of $M$, we need 
to study only the sub-system projected onto the surface $M$. For this we define a new set 
of two variables induced on $M$. In particular, this surface is defined by $2 X_2^2+2 X_2 
X_3+X_3^2= 1$, hence defining $X_2=\cos\theta$ 
and $X_3=\sin\theta-\cos\theta$ we have our first variable, namely $\theta$, while the 
second will be $Y_1$. Since on $M,$ we have that $X_2\geq 0$, and additionally, $X_2^2+X_3^2+Y_
1^2\leq 1$ we get $\theta \in\left[\cos ^{-1}\left(\frac{2}{\sqrt{5}}\right),\frac{\pi}{2}\right]$. The system of equation on the surface $M$ then becomes
\begin{eqnarray}
\tilde \theta &=-\left(\frac{\cos\theta - \sin\theta}{2}\right)  \left[1 + 2 Y_1^2 + \cos(2 
\theta) - 2 \sin(2 
\theta)\right]
\label{eqexp1}\\
\tilde Y_1 &=Y_1\left( \frac{\cos\theta + \sin\theta}{2} \right) \left[1 + 2 Y_1^2 + \cos(2 
\theta) - 2 \sin(2 
\theta)\right],
\label{eqexp2}
\end{eqnarray}
whose orbits are trivially given by 
\begin{eqnarray}
Y_1(\theta )= c_1 \left[\cos (\theta )-\sin (\theta )\right].
\end{eqnarray} 
In order to make the behavior of this case more transparent, in Fig. \ref{Figexp}  we 
depict some orbits of the system (\ref{eqexp1})-(\ref{eqexp2}) representing the dynamics 
on the invariant surface $M$. As we can see, the Minkowski point $P_M$ is the local sink 
on this surface, while has we have already mentioned it, it is a saddle point in the 3D phase space. Finally, the dynamics on the invariant set $\bar M$ is the reversed of 
this figure.

In summary, we deduce that we have the same conclusions previously encountered in the 
massless case. The solution connecting the horizon ($P_H$) to an asymptotic flat region ($P_M$) is 
unique and it is the Schwarzschild solution (green curve in Fig.\ref{Figexp}). All other solutions exhibit naked singularities. This is consistent with the no-go theorem, which states that for any 
convex potential $(V_{,\psi\psi}>0)$ , the Schwarzschild spacetime
is the unique static black hole solution which is asymptotically flat.

For the second  case, namely when the trajectory flows from one surface to the other ($M$ 
to $\bar M$),  we define a negative definite function:
\begin{eqnarray}
h=\frac{X_2+X_3+\sqrt{1-X_2^2}}{X_2+X_3-\sqrt{1-X_2^2}},
\end{eqnarray}
and we obtain $\tilde h=-A h$, where
\begin{eqnarray}
A=\frac{2 \left(Y_1^2+1\right)-(X_2+X_3) \left(2 X_2+\sqrt{2} \lambda  
Y_1\right)}{\sqrt{1-X_2^2}}.
\end{eqnarray} 
The function $h$ is defined as the ratio between the equations defining $\bar M$ 
$\Bigl(X_2+X_3+\sqrt{1-X_2^2}=0\Bigr)$ and $M$ $\Bigl(X_2+X_3-\sqrt{1-X_2^2}=0\Bigr)$. 
Hence, for the points satisfying $A>0$ the orbits transit from $M$ to $\bar{M}$. Using 
the coordinates of the critical points, it is easy to check that $A>0$ for the critical 
points $P_H$,$P_M$ and $P_S$. Therefore, any solution starting near one of these critical 
points will evolve to the surface $\bar M$. The other points located on the critical 
curve (the red curve in Fig. \ref{Fig:exp3D}), are repelling for some values of 
$\lambda$ and saddle elsewhere, however they are never stable. Thus, any 
trajectory starting near $M$ will end in the hypersurface $\bar 
M$. Moreover, we found an additional critical point which appears for $\lambda^2>6$
and it is unstable, while its dual in the region $x_2<0$ is stable. These 
additional points will also generate solutions from the region $x_2>0$ to the region 
$x_2<0$. Hence, all these solutions will cross the surface $x_2=0$, nevertheless  by 
violating the flare-out condition $(\tilde x_2=-y_1^2<0)$, which implies a  maximum radius at 
$x_2=0$. These solutions are the exact opposite of wormhole solutions encountered 
previously. 

We can now perform an expansion around each critical point, in order to reveal the 
behavior of the metric around it. For instance, for the two additional 
critical 
points $P_\lambda:=\left(\frac{\sqrt{2}}{\sqrt{\lambda ^2+4}},\frac{\sqrt{2}}{\sqrt{\lambda ^2+4}},\frac{\lambda }{\sqrt{\lambda ^2+4}}\right)$, $\bar P_\lambda:= \left(-\frac{\sqrt{2}}{\sqrt{\lambda ^2+4}},-\frac{\sqrt{2}}{\sqrt{\lambda ^2+4}},-\frac{\lambda }{\sqrt{\lambda ^2+4}}\right)$, which exist only when $\lambda^2>6$ (otherwise the scalar field will be phantom),  we have
\begin{eqnarray}
\label{Plambda}
A\propto r^2,\quad B\propto r^{2-\lambda^2},
\end{eqnarray}
and since $B$ diverges ($x_2\rightarrow\infty$) we approach these points in the limit 
$r\rightarrow 0$. This is consistent with the global picture we had previously: 
the solution starts from a minimum radius (at $r=0$), it evolves until a maximum radius 
($x_2=0$), and then the radius decreases again to $r=0$.
The same analysis can be performed for the critical curves, for which we found
\begin{eqnarray}
A\propto r^{2\bigl(-1+\frac{\sqrt{1-X_2^2}}{|X_2|}\bigr)},\quad B\propto r^{-2 
\frac{\sqrt{1-X_2^2}}
{|X_2|}},
\end{eqnarray}
where $X_2$ is the coordinate of each point on the critical curves 
($-2/\sqrt{5}<X_2<2/\sqrt{5}$). We again see that the critical curves are approached when 
$r\rightarrow 0$ (i.e $B\rightarrow \infty$). Hence, each solution connects a singularity 
at $r=0$ ($x_2>0$) to another singularity at $r=0$ ($x_2<0$), through a maximum radius at 
$x_2=0$. Additionally, we have solutions connecting a non singular point, such as the 
horizon $P_H$ or the Minkowski space $P_M$, to a singular space at $x_2<0$. Finally, we 
have solutions connecting a singular space (red  critical curve in Fig. \ref{Fig:exp3D} 
for $x_2>0$) to a horizon $\bar P_H$, which is also a naked singularity. We therefore see
that, for the exponential potential, there is no other regular solution apart from 
the Schwarzschild one.

Finally we mention that if we consider the case where the potential is negative, then we 
have to reverse the allowed phase space. Specifically, only the regions in blue in   
Fig. \ref{Fig:exp3D} are accessible. Hence, we have two disconnected phase 
spaces. We have the same critical points as described previously  for the positive 
potential case (see Fig. \ref{Fig:exp3D}), however the points 
$P_\lambda$ and $\bar P_\lambda$ are now located in the blue region for 
$\lambda^2<6$. In this new phase space the value $\lambda=0$ brings an additional critical 
point (since now $\lambda^2<6$) and it corresponds to a constant potential $V(\phi)=V_0 e^{0 \phi}=V_0$. According to 
Eq. (\ref{Plambda})  this additional point is an AdS solution, and it is reached when 
$r\rightarrow \infty$. But this critical point corresponds to a constant potential and not exponential and therefore can't be considered in this analysis. We consider for the exponential case only $\lambda\neq 0$. Therefore, we conclude that AdS critical point does not exist for the exponential potential, and therefore there are no solutions with an exponential potential asymptotically AdS.
\section{On some aspects of the general case}
\label{Generalcase}

Finally, we close this work by trying to extract information for the  case  of a general 
potential. In this case, the only critical points at the finite region of the phase 
space 
\eqref{phase_space_gen}, (for positive potential)  correspond to Minkowski spacetime,  since any other point will be 
excluded due to the reality condition of the scalar field. In particular, the critical points are  
\begin{enumerate}
	\item[$P_M$:] $(x_2=1, x_3=0,y_1=0,\lambda=\lambda_c)$. The corresponding 
eigenvalues are 
$2,-1,-1,0$.

	\item[$\bar P_M$:] $(x_2=-1, x_3=0,y_1=0,\lambda=\lambda_c)$. The 
corresponding eigenvalues are 
$-2,1,1,0$.
\end{enumerate}
Hence,
both critical lines are saddle (where $\lambda_c$ is any real number). Therefore, in general, a solution with a potential will 
not be asymptotically flat. But the Minkowski critical line can be stable if the phase space is 
reduced to a stable invariant submanifold containing it. In fact, $P_M$ has 2 attractive directions (2 negative eigenvalues) and therefore Minkowski is an attractor on a 2 dimensional sub-space containing these two attractive directions. In the case where 
$\lambda$ is finite, it is easy to find an invariant submanifold containing Minkowski as a 
critical point, namely
\begin{eqnarray}
x_2^2+2x_2 x_3-y_1^2-1=0.
\end{eqnarray}
According to the condition (\ref{Gauss_2}) we have that $y_2=0$, which implies that this 
subspace corresponds to $V=0$ which we previously studied.
\vspace{0.5cm}

\textit{In summary, we conclude that for any non-negative potential which is not asymptotically zero 
(i.e. corresponding to finite $\lambda$), the unique flat black hole solution is the 
Schwarzschild spacetime.}
\vspace{0.5cm}
\newline
For completeness, we consider also negative potential for which an additional critical point appears. We will study only non-phantom scalar field ($\varepsilon=+1$). In this case, apart from the Minkowski points, $P_M, \bar P_M$, we have the new points  

\begin{enumerate}[align=left]
\item[$Q(\lambda ^\star)$:] $(x_2,x_3,y_1,\lambda)=$ $\left(\frac{{\lambda^\star}^2}{\sqrt{{\lambda^\star}^4-4}}, \frac{2}{\sqrt{{\lambda^\star}^4-4}}, \frac{\sqrt{2} \lambda
   ^\star}{\sqrt{{\lambda^\star}^4-4}}, \lambda ^\star\right)$, 
\item[$\bar Q(\lambda ^\star)$:] $(x_2,x_3,y_1,\lambda)=$  $\left(-\frac{{\lambda^\star}^2}{\sqrt{{\lambda^\star}^4-4}}, -\frac{2}{\sqrt{{\lambda^\star}^4-4}}, -\frac{\sqrt{2} \lambda
   ^\star}{\sqrt{{\lambda^\star}^4-4}}, \lambda ^\star\right)$,
\end{enumerate}
where $\lambda^\star$ are the roots of the function $f$, $f(\lambda ^\star)=0$. These points exist only if ${\lambda ^\star}^2>2$ and $V(\psi)<0$, because $3y_2\equiv V(\psi)/K=1-x_2^2-2x_2x_3+y_1^2<0$. 
\newline
The corresponding eigenvalues of $Q(\lambda ^\star)$ are 
 $-\frac{{\lambda^\star}^2+2}{\sqrt{{\lambda^\star}^4-4}},-\frac{2+{\lambda^\star}^2+\sqrt{-7 {\lambda^\star}^4+4 {\lambda^\star}^2+36}}{2
   \sqrt{{\lambda^\star}^4-4}}$,\\
	$-\frac{2+{\lambda^\star}^2-\sqrt{-7 {\lambda^\star}^4+4 {\lambda^\star}^2+36}}{2 \sqrt{{\lambda
   ^\star}^4-4}},-\frac{2 \lambda ^\star f'\left(\lambda ^\star\right)}{\sqrt{{\lambda^\star}^4-4}}$. For $f'\left(\lambda
   ^\star\right)=0$ the fixed point is non-hyperbolic. On the other hand, for $\lambda ^\star f'\left(\lambda ^\star\right)>0$, the critical point is an attractor while it is saddle otherwise. As previously noticed, $\bar Q(\lambda ^\star)$ is a complementary point and therefore has an exact opposite behavior. Notice that if $\lambda^\star=\infty$, the critical point reduces to the Minkowski spacetime, otherwise it represents a non physical spacetime where the metric is defined by $B=x_2^2={\lambda^\star}^4/({\lambda^\star}^4-4)$ and $A=r^{4/{\lambda^\star}^2}$.

To study the behavior of the orbits at infinity, we use the Poincar\'e variables \eqref{Poincare_exp}  to examine the limit where $x_2^2+x_3^2+y_1^2$ becomes infinity (we assume that $\lambda$ is finite). After this compactification, we found
\begin{subequations}
\label{Poincare_3}
\begin{align}
\tilde X_2=&-\frac{\lambda}{\sqrt{2}} X_2 Y_1\left[2 X_2^2+2 X_2 X_3+X_3^2-1\right] +Y_1^2 \left[X_2 (2X_2+X_3)-1\right] \nonumber\\
&-X_2 X_3 \left[X_2 (3 X_2+X_3)-2\right],\\
\tilde X_3=&-\frac{\lambda}{\sqrt{2}} X_3 Y_1 \left[2 X_2^2+2 X_2 X_3+X_3^2-1\right]-3 X_2^2 X_3^2+2 X_2^2-X_2 X_3^3 \nonumber\\
&+X_3 Y_1^2 (2 X_2+X_3)  +X_2 X_3+X_3^2-1,\\
\tilde Y_1= &\frac{\lambda}{\sqrt{2}}\left[1-Y_1^2\right] \left[2 X_2^2+2 X_2 X_3+X_3^2-1\right] -X_2 Y_1 \left[3 X_2 X_3+X_3^2+1\right]  \nonumber\\
&+Y_1^3 [2 X_2+X_3],\\
\tilde \lambda =& -\sqrt{2} Y_1 f\left(\lambda\right),\label{eq55d}
\end{align}
\end{subequations}
where, as before, we have introduced a new radial variable through $\tilde{X}\rightarrow 
\sqrt{1-X_2^2-X_3^2-Y_1^2}~X'$. The system \eqref{syst44} is recovered for $f(\lambda)=0, \lambda=\text{constant}$, as expected. 
In this case the dimensionality of the dynamical system reduces to 3D. 
\newline
For positive potentials the system \eqref{Poincare_3} defines a flow on the phase space 
\begin{align}
\left\{(X_2,X_3,Y_1,\lambda):2 X_2^2+2 X_2 X_3+X_3^2\leq 1,X_2^2+X_3^2+Y_
1^2\leq 1, \lambda \in \mathbb{R}\right\},
\end{align} 
whereas, for negative potentials $V(\psi)<0$, the phase space is given by 
\begin{align}
\label{eq58}
\left\{(X_2,X_3,Y_1,\lambda):2 X_2^2+2 X_2 X_3+X_3^2\geq 1, X_2^2+X_3^2+Y_
1^2\leq 1, \lambda \in \mathbb{R}\right\}.
\end{align}


\begin{table}
\caption{\label{TableIII} Critical points at infinity for the Poincar\'e (global) system \eqref{Poincare_3}, for the case of a canonical scalar field with arbitrary potential. We have defined the surfaces $\bar M$ $\Bigl(X_2+X_3+\sqrt{1-X_2^2}=0\Bigr)$ and $M$ $\Bigl(X_2+X_3-\sqrt{1-X_2^2}=0\Bigr)$.}
\footnotesize
\begin{adjustwidth}{-2cm}{}
\begin{tabular}{@{}llllllll}
\br
Point & $X_2$ & $X_3$ & $Y_1$ & $\lambda$ & Existence & Stability & Nature\\
\mr 
$P_H$  & $0$& $1$ & $0$ & $\lambda_c$ & $\lambda_c\in\mathbb{R}$ & unstable & Horizon \\
$\bar P_H$  & $0$& $-1$ &$0$ & $\lambda_c$ & $\lambda_c\in\mathbb{R}$ & stable & Horizon\\
$P_S$  & $\frac{2}{\sqrt{5}}$ & $-\frac{1}{\sqrt{5}}$ & $0$ & $\lambda_c$ & $\lambda_c\in\mathbb{R}$ & unstable & Singularity\\
$\bar P_S$ & $-\frac{2}{\sqrt{5}}$ & $\frac{1}{\sqrt{5}}$ & $0$  & $\lambda_c$ & $\lambda_c\in\mathbb{R}$ & stable & Singularity\\
$P_M$ & $\frac{1}{\sqrt{2}}$ & $0$ & $0$  & $\lambda_c$ & $\lambda_c\in\mathbb{R}$  & saddle &  Minkowski\\
$\bar P_M$ & $-\frac{1}{\sqrt{2}}$ & $0$ & $0$  & $\lambda_c$ & $\lambda_c\in\mathbb{R}$  & saddle &  Minkowski\\
$P_{AdS}$ & $\frac{1}{\sqrt{2}}$ & $\frac{1}{\sqrt{2}}$ & $0$  & $0$ & $V(\psi)<0$ &  stable for $f(0) > 0$ & Anti-de Sitter\\
$\bar P_{AdS}$ & $-\frac{1}{\sqrt{2}}$ & $-\frac{1}{\sqrt{2}}$ & $0$  & $0$ & $V(\psi)<0$ & unstable for $f(0) > 0$ & Anti-de Sitter\\

$Q(\lambda^\star)$ & $\frac{{\lambda^\star}^2}{\sqrt{2 {\lambda^\star}^2(1+{\lambda^\star}^2)}}$ & $\frac{\sqrt{2}}{\sqrt{{\lambda^\star}^2(1+{\lambda^\star}^2)}}$ & $\frac{{\lambda^\star}}{\sqrt{{\lambda^\star}^2(1+{\lambda^\star}^2)}}$  & $\lambda ^\star$ & $V(\psi)<0$, & stable for ${\lambda^\star}f'\left(\lambda^\star\right)>0$& $A\propto r^{4/{\lambda^\star}^2}$, \\
			&&&&& $f(\lambda ^\star)=0$ & saddle otherwise & $B\propto {\lambda^\star}^4/({\lambda^\star}^4-4)$ \\ 
			&&&&& ${\lambda^\star}^2\geq 2$ &  &  \\ 

$\bar Q(\lambda^\star)$ & $-\frac{{\lambda^\star}^2}{\sqrt{2 {\lambda^\star}^2(1+{\lambda^\star}^2)}}$ & $-\frac{\sqrt{2}}{\sqrt{{\lambda^\star}^2(1+{\lambda^\star}^2)}}$ & $-\frac{{\lambda^\star}}{\sqrt{{\lambda^\star}^2(1+{\lambda^\star}^2)}}$  & $\lambda ^\star$ & $V(\psi)<0$, & unstable for ${\lambda^\star}f'\left(\lambda^\star\right)>0$& $A\propto r^{4/{\lambda^\star}^2}$, \\
			&&&&& $f(\lambda ^\star)=0$ & saddle otherwise & $B\propto {\lambda^\star}^4/({\lambda^\star}^4-4)$ \\ 
			&&&&& ${\lambda^\star}^2\geq 2$ &  &  \\ 

$P(\lambda^\star)$ & $\sqrt{\frac{2}{{\lambda ^\star}^2+4}}$ & $\sqrt{\frac{2}{{\lambda ^\star}^2+4}}$ & $\frac{\lambda ^\star}{\sqrt{{\lambda ^\star}^2+4}}$  & $\lambda ^\star$ &$f(\lambda ^\star)=0$, & unstable for & $A\propto r^2, B\propto r^{2-(\lambda^\star)^2}$ \\
			&&&&& ${\lambda ^\star}^2\geq 6$ & $f'\left(\lambda^\star\right)<0, \lambda^\star>\sqrt{6}$, & singularity \\ &&&&&&  or $f'\left(\lambda^\star\right)>0,
\lambda^\star<-\sqrt{6}$& \\
&&&&&& saddle otherwise &\\
$\bar P(\lambda^\star)$ & $-\sqrt{\frac{2}{{\lambda ^\star}^2+4}}$ & $-\sqrt{\frac{2}{{\lambda ^\star}^2+4}}$ & $-\frac{\lambda ^\star}{\sqrt{{\lambda ^\star}^2+4}}$  & $\lambda ^\star$&$f(\lambda ^\star)=0$, & stable for & $A\propto r^2, B\propto r^{2-(\lambda^\star)^2}$ \\
			&&&&& ${\lambda ^\star}^2\geq 6$ & $f'\left(\lambda^\star\right)<0, \lambda^\star>\sqrt{6}$, & singularity \\ &&&&&&  or $f'\left(\lambda^\star\right)>0,
\lambda^\star<-\sqrt{6}$& \\
&&&&&& saddle otherwise &\\
$\mathcal{C}(\lambda^\star)$ & $\cos\theta$ & $\sin\theta-\cos\theta$ & $\sqrt{1-X_2^2-X_3^2}$  & $\lambda ^\star$ &$f(\lambda ^\star)=0$ & $M$ is unstable for  & singularity \\ 
&&&&&  & $f'\left(\lambda^\star\right)<0$, & \\
&&&&&  & $\lambda^\star<\frac{2\cos(\theta)+2\sin(\theta)}{\sqrt{-1-\cos(2\theta)+2\sin(2\theta)}}$& \\
$$ & $\cos\theta$ & $\sin\theta-\cos\theta$ & $-\sqrt{1-X_2^2-X_3^2}$  & $\lambda ^\star$ &$f(\lambda ^\star)=0$ & $M$ is unstable for  & singularity \\ 
&&&&&  & $f'\left(\lambda^\star\right)>0$, & \\
&&&&&  & $\lambda^\star>-\frac{2\cos(\theta)+2\sin(\theta)}{\sqrt{-1-\cos(2\theta)+2\sin(2\theta)}}$& \\
&&&&&& saddle otherwise &\\
$\bar{\mathcal{C}}(\lambda^\star)$ & $\cos\theta$ & $-\sin\theta-\cos\theta$ & $\sqrt{1-X_2^2-X_3^2}$  & $\lambda ^\star$ &$f(\lambda ^\star)=0$ & $M$ is stable for  & singularity \\ 
&&&&&  & $f'\left(\lambda^\star\right)>0$, & \\
&&&&&  & $\lambda^\star>\frac{2\cos(\theta)-2\sin(\theta)}{\sqrt{-1-\cos(2\theta)+2\sin(2\theta)}}$& \\
$$ & $\cos\theta$ & $-\sin\theta-\cos\theta$ & $-\sqrt{1-X_2^2-X_3^2}$  & $\lambda ^\star$ &$f(\lambda ^\star)=0$ & $M$ is stable for  & singularity \\ 
&&&&&  & $f'\left(\lambda^\star\right)<0$, & \\
&&&&&  & $\lambda^\star<-\frac{2\cos(\theta)-2\sin(\theta)}{\sqrt{-1-\cos(2\theta)+2\sin(2\theta)}}$& \\
&&&&&& saddle otherwise &\\
\br
\end{tabular}\\
\end{adjustwidth}
\end{table}
\normalsize


In the following list are enumerated half of the critical points/lines at infinity $(X_2,X_3,Y_1,\lambda)$ of the system \eqref{Poincare_3}, since the points with barred labels and the corresponding unbarred ones have the opposite dynamical behavior (see the summary in table \ref{TableIII}): 
\begin{enumerate}[align=left]
\item[$P_H$:] $\left(0, 1, 0,\lambda_c\right)$ with $\lambda_c\in\mathbb{R}$. The eigenvalues are $2,2,0,0$, thus the point is nonhyperbolic. The center manifold of $P_H$ is the 2-dimensional surface defined by the boundary of $M$; $X_2+X_3-\sqrt{1-X_2^2}=0, X_2^2+X_3^2+Y_1^2=1$, $\lambda\in\mathbb{R}$ and the critical point is unstable (see the full analysis in \ref{centerPH}). This point corresponds to the horizon as analyzed previously.

\item[$P_S$:] $\left(\frac{2}{\sqrt{5}},-\frac{1}{\sqrt{5}},0,\lambda_c\right)$ with $\lambda_c\in\mathbb{R}$. The eigenvalues are $\frac{6}{\sqrt{5}},\frac{2}{\sqrt{5}},0,0$, thus the point is nonhyperbolic. The center manifold of $P_S$ is the same as for $P_H$ and it is also unstable (see the full analysis in \ref{centerPS}). As seen, in the massless case, this point corresponds to a singularity where $A(r)=B(r)\simeq r^{-1}$.
 
\item[$ P(\lambda^\star)$:] $\left(\sqrt{\frac{2}{{\lambda ^\star}^2+4}},\sqrt{\frac{2}{{\lambda ^\star}^2+4}},\frac{\lambda ^\star}{\sqrt{{\lambda ^\star}^2+4}},\lambda ^\star\right)$. It exists for $f(\lambda ^\star)=0, {\lambda ^\star}^2\geq 6$. The eigenvalues are $\frac{{\lambda ^\star}^2-6}{\sqrt{2({\lambda ^\star}^2+4)}},\frac{{\lambda ^\star}^2-6}{\sqrt{2({\lambda ^\star}^2+4)}},\frac{\sqrt{2}
   \left({\lambda ^\star}^2-2\right)}{\sqrt{{\lambda ^\star}^2+4}},-\frac{\sqrt{2} \lambda ^\star f'\left(\lambda ^\star\right)}{\sqrt{{\lambda ^\star}^2+4}}$. It is unstable for $f'\left(\lambda^\star\right)<0, \lambda^\star>\sqrt{6}$, or $f'\left(\lambda^\star\right)>0,
\lambda^\star<-\sqrt{6}$, and saddle otherwise. This point has already been studied in the case of an exponential potential.

\item[$C(\lambda^\star)$:] is given by the conditions $f(\lambda ^\star)=0$, $2X_2^2+2X_2X_3+X_3^2=1$, $X_2^2+X_3^2+Y_1^2=1$. This line has already been studied in the exponential potential case where it was represented by the red curve in Fig.\ref{Fig:exp3D}. It is the boundary of $M$ and exists only if $f(\lambda)$ has roots. Notice that for a given root of $f(\lambda)$, this line corresponds to the center manifold of $P_H$ and $P_S$. By using the same analysis than the exponential case, we found that the line is unstable for $f'\left(\lambda ^\star\right)<0, \lambda ^\star<\frac{2\cos{\theta}+2\sin\theta}{\sqrt{-1-\cos(2\theta)+2\sin(2\theta)}}$ and $Y_1>0$ or $f'\left(\lambda ^\star\right)>0, \lambda ^\star>-\frac{2\cos{\theta}+2\sin\theta}{\sqrt{-1-\cos(2\theta)+2\sin(2\theta)}}$ and $Y_1<0$ while $\arccos(2/\sqrt{5})<\theta<\pi/2$.

\item[$P_{AdS}$:] $\left(\frac{1}{\sqrt{2}},\frac{1}{\sqrt{2}},0,0\right)$. This point belongs to the phase space \eqref{eq58}, that is, it exists only for negative potential $V(\psi)<0$. The eigenvalues are $-\frac{3}{\sqrt{2}},-\sqrt{2},-\frac{\sqrt{9-12 f(0)}+3}{2 \sqrt{2}},-\frac{3-\sqrt{9-12 f(0)}}{2 \sqrt{2}}$. It is a sink for $f(0)>0$ and corresponds to AdS spacetime. Thus, the solutions are asymptotically AdS if we impose the condition $V(\psi)<0$ and $f(0)>0$ (see \ref{PAdS} for complete analysis). 
\end{enumerate} 
This last relation is very useful to check if a model has solutions which are asymptotically AdS (AAdS). For example the Martinez-Troncoso-Zanelli (MTZ) black hole \cite{Martinez:2004nb} is AAdS. For this model the potential is defined as
\begin{eqnarray}
V(\psi)=\Lambda\Bigl[1+2\sinh\Bigl(\frac{\psi}{\sqrt{6}}\Bigr)\Bigr]\,,~\Lambda<0
\end{eqnarray}
It is easy to check that for this potential $f(\lambda)=2/3-\lambda^2$ and therefore $f(0)>0$, which implies that an AAdS solution exists, which is the MTZ black hole.
\newline
\newline
Finally, it is interesting to expand this analysis to asymptotically flat potentials for which $\lambda$ can diverge.

\subsection{Analysis as $\lambda\rightarrow +\infty$}

In this section, we investigate only the limit $\lambda\rightarrow +\infty$, since the limit $\lambda\rightarrow -\infty$ can be studied through the change $\lambda \rightarrow -\lambda$ in the analysis to follow. So far, we have shown that in the phase space, Minkowski spacetime appears as a critical line. We have shown that any trajectory connecting to a Minkowski critical point for any finite value of $\lambda$, is trivial (Schwarzschild) or a singular spacetime. We would like to study in this section, orbits connecting to the Minkowski critical point localized at infinity in the phase space $\lambda=\infty$.

For that, we consider the coordinate transformation $v=g(\lambda)=\lambda^{-1}$ which maps the interval $[\lambda_0, +\infty[$ onto $]0,\epsilon]$, where $\epsilon=g(\lambda_0)=\frac{1}{\lambda_0}$ and $\lambda_0$ is any positive real number. 
The system \eqref{Poincare_3} is then topologically equivalent to the system
\begin{subequations}
\label{infinity_1}
\begin{align}
\frac{d X_2}{d \varsigma}=&v \left[Y_1^2 (X_2 (2 X_2+X_3)-1) -X_2 X_3 (X_2 (3 X_2+X_3)-2)\right] \nonumber\\
&-\frac{ X_2 Y_1  \left(2 X_2^2+2 X_2 X_3+X_3^2-1\right)}{\sqrt{2}},\\
\frac{d X_3}{d \varsigma}=&v \left[X_2^2 \left(2-3 X_3^2\right)+X_2 \left(-X_3^3+2 X_3
   Y_1^2+X_3\right) +X_3^2 \left(Y_1^2+1\right)-1\right] \nonumber\\
  & -\frac{ X_3 Y_1  \left(2 X_2^2+2 X_2X_3+X_3^2-1\right)}{\sqrt{2}},\\
\frac{d Y_1}{d \varsigma}=&v \left(Y_1^3 (2 X_2+X_3)-X_2 Y_1 \left(3 X_2 X_3+X_3^2+1\right)\right)\nonumber\\
&-\frac{\left(Y_1^2-1\right)  \left(2 X_2^2+2 X_2 X_3+X_3^2-1\right)}{\sqrt{2}},\\
\frac{d v}{d \varsigma}=&\sqrt{2} v^3 Y_1 f\left(\frac{1}{v}\right),
\end{align}
\end{subequations}
where we have introduced the derivative $\frac{d f}{d\varsigma}=g(\lambda)\tilde{f}$.
\newline
If we consider positive potentials, $V(\psi)\geq 0$, the above system defines a flow on the phase space 
\begin{align}
\label{eq60}
\left\{(X_2,X_3,Y_1,v):2 X_2^2+2 X_2 X_3+X_3^2\leq 1, X_2^2+X_3^2+Y_1^2\leq 1,0\leq v \leq \epsilon\right\}, \epsilon\ll 1.
\end{align}
 
The fixed points (or fixed surfaces) of the system \eqref{infinity_1} at infinity (i.e., satisfying $v=0$ and $v^3f(1/v)=0$) on the phase space \eqref{eq60} are:
\begin{itemize}[align=left]
\item[$M,\bar M$:] $2X_2^2+2X_2X_3+X_3^2=1, v=0$. These surfaces are the two boundaries encountered in the exponential case and they correspond to $V(\psi)=0$. The eigenvalues are $0,0,-\sqrt{2} Y_1,\sqrt{2} Y_1 (3 \beta_{1}-\beta_{2})$, where $\beta_1=\lim_{v\rightarrow 0} v^2 f\left(\frac{1}{v}\right), \beta_2=\lim_{v\rightarrow 0} v f'\left(\frac{1}{v}\right).$ These surfaces are saddle for $3 \beta_{1}-\beta_{2}>0$. For $3 \beta_{1}-\beta_{2}<0$ and depending on the sign of $Y_1$, the surface will be a 2D unstable manifold (respectively, stable manifold) for $Y_1<0$ (respectively, $Y_1>0$).  
  
   
\item[$P(\infty)$:] $X_2=0, X_3=0, Y_1=1, v=0$. It is the endpoint of the line $P(\lambda):=\left(\frac{\sqrt{2}}{\sqrt{\lambda ^2+4}},\frac{\sqrt{2}}{\sqrt{\lambda ^2+4}},\frac{\lambda }{\sqrt{\lambda ^2+4}},\frac{1}{\lambda }\right)$ as $\lambda\rightarrow \infty$. The eigenvalues are $\sqrt{2},\frac{1}{\sqrt{2}},\frac{1}{\sqrt{2}},\sqrt{2} (3 \beta_{1}-\beta_{2}).$ This point is unstable if $3 \beta_{1}-\beta_{2}>0$, or saddle otherwise.
\item[$\bar P(\infty)$:] $X_2=0, X_3=0, Y_1=-1, v=0$. It is the endpoint of the line $\bar P(\lambda):=\left(-\frac{\sqrt{2}}{\sqrt{\lambda ^2+4}},-\frac{\sqrt{2}}{\sqrt{\lambda ^2+4}},-\frac{\lambda }{\sqrt{\lambda ^2+4}},\frac{1}{\lambda }\right)$ as $\lambda\rightarrow \infty$. The eigenvalues are $-\sqrt{2}, -\frac{1}{\sqrt{2}}, -\frac{1}{\sqrt{2}}, -\sqrt{2} (3 \beta_{1}-\beta_{2}).$  This point is stable if $3 \beta_{1}-\beta_{2}>0$, or saddle otherwise.
\end{itemize}
It is worth noticing that the stability behavior of the points for $\lambda\rightarrow +\infty$ depends crucially on the limit  
\begin{eqnarray}
\lim_{\lambda\rightarrow \infty} \frac{3 f(\lambda)}{\lambda^2}-\frac{f'(\lambda)}{\lambda}=\lim_{\lambda\rightarrow \infty} \left\{(\Gamma(\lambda)-1)-\lambda \Gamma'(\lambda)\right\}\equiv3\beta_1-\beta_2\nonumber
\end{eqnarray}
While the points $P(\infty)$ and $\bar P(\infty)$ correspond to singular points and therefore are not physically interesting, the points on the critical surfaces $M$ and $\bar M$ with coordinates $(X_2=\pm\frac{1}{\sqrt{2}},X_3=0,Y_1=0,v=0)$ correspond to Minkowski spacetime. Thus, to have a black hole solution asymptotically flat, an orbit should connect to one of these points corresponding to Minkowski spacetime. In conclusion, any asymptotically flat black hole solution needs to connect the horizon located at ($X_2=0,X_3=1,Y_1=0,\lambda$), where $\lambda$ is any real number, to the Minkowski spacetime ($X_2=\frac{1}{\sqrt{2}},X_3=0,Y_1=0,\lambda=\infty$). Therefore, two variables diverge along this orbit, $x_3$ near the horizon and $\lambda$ asymptotically.

%

For example, in \cite{Anabalon:2012ih} the authors found a potential which gives an asymptotically flat solution. We found for this potential that the orbits connect the horizon to the Minkowski spacetime at infinity in the phase space $\lambda=\infty$. Along this orbit, we have verified that the two variables $(x_3,\lambda)$ diverge in accordance with our result.

Additionally, considering the potential studied numerically in \cite{Nucamendi:1995ex}
\begin{align}
V(\psi)=\frac{\alpha}{4}    \Bigl[(\psi-a)^2 -\frac{4(\eta_1+\eta_2)}{3} (\psi-a)+2 \eta_1 \eta_2 \Bigr] (\psi-a)^2,
\end{align}
we found the same result. Therefore, we see that it is fundamental to study completely the phase space at infinity. A simple Poincar\'e compactification gives non-hyperbolic critical points and therefore their study by standard tools is impossible. We postpone this analysis to a future project where the goal will be to find suitable variables in order to have hyperbolic critical points.

We would like to end this section with a general comment on asymptotically flat spacetimes. In fact, we have seen in this section that we can have asymptotically flat spacetimes ``localized'' at finite $\lambda$ or at $\lambda=\infty$. We see therefore that we can have e.g. a Minkowski spacetime with finite $\lambda$ and therefore with a particular value of the scalar field and another Minkowski spacetime with infinite $\lambda$ which would correspond to another value of the scalar field. If they usually correspond to two different solutions, they can be connected in the phantom case through a throat at $x_2=0$ which will correspond to a wormhole solution connecting two asymptotically flat spacetimes but corresponding to two different values of the scalar field. For example, this could happen when the scalar field has a symmetry breaking potential with two minima. One minimum will correspond to finite potential and $V'(\phi)=0$ (extremum of the potential) which would imply $\lambda=0$ and another minimum corresponding to $V(\phi)=0$ and therefore $\lambda=\infty$. These situations can be encountered in a Higgs-like potential $V(\phi)=a(\phi^2-b)^2$. These solutions are very interesting because they connect two vacuum states corresponding to the minima of potential which corresponds to a kink-like configuration of the scalar field where the scalar field is equal to some value in one asymptotically flat region and another one in the other region, varying substantially near the throat. An interesting example, of such case is described by the Kodama's solution \cite{Kodama:1978dw}.

\section{Conclusions}
\label{Conclusions}

In this work, we reformulated some of the important results on black holes in the
presence of a minimally coupled scalar field, by using the dynamical system analysis. 
Using the $1+1+2$ formalism in the case of spherically symmetric spacetimes and 
suitable normalized variables with the help of the Gaussian curvature, we could 
reformulate the Einstein equations as first order differential equations. These equations 
can be investigated using the techniques of dynamical systems.

We mention that, although in the majority of works in the literature one 
suitably reconstructs the scalar potential in order to find a black hole solution, in the 
present work we follow the more physical approach to define the potential from the 
start since in any realistic theory the potential is given by the theory a priori.

As a first case, we examined the zero potential, i.e. the case of a massless scalar 
field. We were able to reduce the phase space to 2D, and then to study its global 
behavior. We recovered all known black hole results, and we found that  apart from the 
Schwarzschild solution all other solutions are naked singularities. This is the well known 
Fisher solution. Additionally, we identified the symmetric phase space which corresponds 
to the white hole part of the solution. Finally, in the case of a phantom field, we were 
able to extract the conditions for the existence of wormholes and analyze the full spectrum of solutions, by defining three types of orbits, which represent three class of solutions, known as Cold Black holes, singular spacetimes and wormholes. The Misner-Sharp mass turns-out to be important in order to distinguish these solutions. The very simple case where $x_3=0$ could be integrated exactly and gives the Ellis wormhole.

In the case of an exponential potential we found that black hole solution which is 
asymptotically flat is unique and it is the Schwarzschild spacetime, while all other 
solutions are naked singularities. Moreover, we found other solution subclasses, which 
connect the two regions of the phase space through $x_2 = 0$ as a wormhole, however by 
violating the flare-out condition, which implies a maximum radius instead of a throat. 
Nevertheless, all these solutions involve naked singularities since they connect 
two singularities or a non-singular to a singular spacetime. Various types of solutions were discussed depending on the value of the parameter $\lambda$.

Finally, generic results have been derived. First we found that in order to have an AAdS spacetime, the potential should be negative and $f(0)>0$. Also, for any potential which is not asymptotically 
zero, $\lambda$ finite, the unique black hole solution is the Schwarzschild spacetime. 
Expanding the analysis to potentials for which $\lambda$ can diverge, we have shown that the only possibility to have a non trivial black hole solution is to have an orbit in the phase space connecting two points at infinity: the horizon ($x_2=0,x_3=\infty,y_1=0,\lambda$), where $\lambda$ is any real number, to the Minkowski spacetime ($x_2=1,x_3=0,y_1=0,\lambda=\infty$). The full analysis of this phase space at infinity will be performed in a future project where the objective is to find conditions on the function $f(\lambda)$ and therefore on the form of the potential in order to have a non trivial black hole solution.

\ack{
R. Gannouji would like to thank Julio Oliva for helpful discussions.
M. Cruz acknowledges CONACyT-M\'exico for support through the grants: Repatriaciones 2015-04 and SNI. 
R. Gannouji thanks DII-PUCV for support through the project No. 039.370/2016. 
A. Ganguly wants to thank NRF (South Africa) and Claude Leon Foundation for financial support. This article is based upon work from COST Action ``Cosmology and Astrophysics Network for Theoretical Advances and Training Actions'', supported by COST (European Cooperation in Science and Technology). 
G. Leon was partially supported by FONDECYT No. 3140244 and acknowldeges DI-VRIEA for financial support through Proyectos VRIEA Investigador Joven 2016 and Investigador Joven 2017.
}
 
\newpage

\appendix

\section{Center Manifold Theory}\label{sectionCM}
In this section, we summarize how to build the center manifold in order to study critical points for which eigenvalues have a null real part. The procedure leads to a reduction of the dimension of the system. For that, we follow the approach in chapter 18 of \cite{wiggins}.\\
We consider vector fields in the form
\begin{eqnarray}
\label{basiceqs}
&\mathbf{x}'=\mathbf{A x}+\mathbf{f}(\mathbf{x},\mathbf{y}),\nonumber\\
&\mathbf{y}'=\mathbf{B y}+\mathbf{g}(\mathbf{x},\mathbf{y}),  \;
(\mathbf{x},\mathbf{y})\in
\mathbb{R}^c\times\mathbb{R}^s,\\
& \mathbf{f(0,0)}=\mathbf{0},
D\mathbf{f(0,0)}=\mathbf{0},\nonumber\\&\mathbf{g(0,0)}=\mathbf{0},
D\mathbf{g(0,0)}=\mathbf{0},\end{eqnarray} where all the eigenvalues of the $c\times c$  matrix $\mathbf{A}$ have zero real parts 
and all eigenvalues of the $s\times s$ matrix $\mathbf{B}$ have negative real parts. The functions $\mathbf{f}$ and $\mathbf{g}$ are $C^r$
functions ($r\geq 2$) with $D\mathbf{f}$ and $D\mathbf{g}$ being their Jacobian matrices respectively.

\begin{defn}[Center Manifold]\label{CMdef}
An invariant manifold will be called a center manifold for
\eqref{basiceqs} if it can locally be represented as follows
\begin{eqnarray}
\label{equationA3}
W^{c}\left(\mathbf{0}\right)  =\left\{  \left(
\mathbf{x},\mathbf{y}\right)
\in\mathbb{R}^c\times\mathbb{R}^s:\mathbf{y}=\mathbf{h}\left(
\mathbf{x}\right) ,\;\left\vert \mathbf{x}\right\vert
<\delta\right\},
\end{eqnarray}
with 
\begin{eqnarray}
\mathbf{h}\left( \mathbf{0}\right)
=\mathbf{0},\;D\mathbf{h}\left( \mathbf{0}\right)  =\mathbf{0},
\end{eqnarray}
for $\delta$ sufficiently small (cf. \cite{wiggins} p. 246,
\cite{Perko},p. 155).
\end{defn}

The conditions $\mathbf{h}\left( \mathbf{0}\right)
=\mathbf{0},\;D\mathbf{h}\left( \mathbf{0}\right)  =\mathbf{0}$
imply that $W^{c}\left(  \mathbf{0}\right)$ is tangent  to $E^c$
at $\left(\mathbf{x},\mathbf{y}\right)=(\mathbf{0},\mathbf{0}),$
where $E^c$ is the generalized eigenspace whose corresponding
eigenvalues have zero real parts. The following three theorems
(see theorems 18.1.2, 18.1.3 and 18.1.4  in \cite{wiggins} p.
245-248) are the main results to the treatment of center
manifolds. The first two are existence and stability theorems of
the center manifold for \eqref{basiceqs} at the origin. The third
theorem allows to compute the center manifold to any desired
degree of accuracy by using Taylor series to solve a quasilinear
partial differential equation that $\mathbf{h}\left(
\mathbf{x}\right)$ must satisfy. The proofs of these results are
given in \cite{Carr:1981}.

\begin{thm}[Existence]\label{existenceCM}
There exists a $C^r$ center manifold for \eqref{basiceqs}. The
dynamics of \eqref{basiceqs} restricted to the center manifold is,
for $\mathbf{u}$ sufficiently small, given by the following
c-dimensional vector field \be\label{vectorfieldCM}
\mathbf{u}'=\mathbf{A
u}+\mathbf{f}\left(\mathbf{u},\mathbf{h}\left(\mathbf{u}\right)\right),\;
\mathbf{u}\in\mathbb{R}^c. \ee
\end{thm}

The next results implies that the dynamics of
\eqref{vectorfieldCM} near $\mathbf{u}=0$ determine the dynamics
of \eqref{basiceqs} near
$\left(\mathbf{x},\mathbf{y}\right)=(\mathbf{0},\mathbf{0})$ (see
also Theorem 3.2.2 in \cite{Guckenheimer}).

\begin{thm}[Stability]\label{stabilityCM}
i) Suppose the zero solution of \eqref{vectorfieldCM} is stable
(asymptotically stable) (unstable); then the zero solution of
\eqref{basiceqs} is also stable (asymptotically stable)
(unstable). Then if $(\mathbf{x}(\tau),\mathbf{y}(\tau))$ is a
solution of \eqref{basiceqs} with $(\mathbf{x}(0),\mathbf{y}(0))$
sufficiently small, then there is a solution $\mathbf{u}(\tau)$ of
\eqref{vectorfieldCM} such that, as $\tau\rightarrow\infty$
\begin{eqnarray}
& \mathbf{x}(\tau)=\mathbf{u}(\tau)+{\cal O}(e^{-r \tau}),\\
& \mathbf{y}(\tau)=\mathbf{h}\left(\mathbf{u}(\tau)\right)+{\cal
O}(e^{-r \tau}),
\end{eqnarray}
where $r>0$ is a constant.
\end{thm}
This theorem says that for initial conditions of
the \emph{full system} sufficiently close to the origin,
trajectories through them asymptotically approach a trajectory on
the center manifold. In particular, singular points sufficiently
close to the origin, sufficiently small amplitude periodic orbits
as well as small homoclinic and heteroclinic orbits are contained
in the center manifold.

To compute the center manifold we proceed as follows: 
suppose we have a center manifold $W^{c}\left(
\mathbf{0}\right)$ defined by \eqref{equationA3}; using the invariance of $W^{c}\left(  \mathbf{0}\right)$ under the dynamics of \eqref{basiceqs}, we derive a quasilinear partial differential equation that $\mathbf{h}\left( \bf{x}\right)$ must satisfy. This is done as follows:

\begin{enumerate}
\item The $(\mathbf{x},\mathbf{y})$ coordinates of any point on
$W^{c}\left(  \mathbf{0}\right) $ must satisfy
\begin{eqnarray}
\mathbf{y}=\mathbf{h}(\mathbf{x}).\label{coordCM}
\end{eqnarray}
\item
Differentiating \eqref{coordCM} with respect to time (or any coordinate of our dynamical system) implies that
the $(\mathbf{x}',\mathbf{y}')$ coordinates of any point on
$W^{c}\left(\mathbf{0}\right) $ must satisfy
\begin{eqnarray}
\mathbf{y}'=D\mathbf{h}\left(\mathbf{x}\right)\mathbf{x}'.\label{totalderivative}
\end{eqnarray}
\item Any
point in $W^{c}\left(  \mathbf{0}\right) $ obey the dynamics
generated by \eqref{basiceqs}. Therefore substituting
\begin{eqnarray}
\mathbf{x}'=\mathbf{A x}+\mathbf{f}\left(\mathbf{x},\mathbf{h}(\mathbf{x})\right),\\
\mathbf{y}'=\mathbf{B}\mathbf{h}(\mathbf{x})+\mathbf{g}\left(\mathbf{x},\mathbf{h}(\mathbf{x})\right),
\end{eqnarray}
into \eqref{totalderivative} gives
\begin{eqnarray}
{\cal N}\left(\mathbf{h}(\mathbf{x})\right)&\equiv
D\mathbf{h}(\mathbf{x})\left[\mathbf{A
x}+\mathbf{f}\left(\mathbf{x},\mathbf{h}(\mathbf{x})\right)\right]
-\mathbf{B}\mathbf{h}(\mathbf{x})-\mathbf{g}\left(\mathbf{x},\mathbf{h}(\mathbf{x})\right) \nonumber\\
&=0.\label{MaineqcM}
\end{eqnarray}
\end{enumerate}

Equation \eqref{MaineqcM} is a quasilinear partial differential
that $\mathbf{h}(\mathbf{x})$ must satisfy in order for its graph
to be an invariant center manifold. To find the center manifold,
all we need to do is solve \eqref{MaineqcM}.

Unfortunately, it is probably more difficult to solve
\eqref{MaineqcM} than our original problem; however the following
theorem gives us a method for computing an approximated solution of
\eqref{MaineqcM} to any desired degree of accuracy.

\begin{thm}[Approximation]\label{approximationCM}
Let $\mathbf{\Phi}:\mathbb{R}^c\rightarrow\mathbb{R}^s$ be a $C^1$
mapping with $\mathbf{\Phi}(\mathbf{0})=\mathbf{0}$ and
$D\mathbf{\Phi}(\mathbf{0})=\mathbf{0}$ such that ${\cal
N}\left(\mathbf{\Phi}(\mathbf{x})\right)={\cal
O}(\|\mathbf{x}\|^q)$ as $\mathbf{x}\rightarrow \mathbf{0}$ for
some $q>1.$ Then,
$|\mathbf{h}(\mathbf{x})-\mathbf{\Phi}(\mathbf{x})|={\cal
O}(\|\mathbf{x}\|^q)$ as $\mathbf{x}\rightarrow \mathbf{0}$.
\end{thm}

This theorem allows us to compute the center manifold to any
desired degree of accuracy by solving \eqref{MaineqcM} to the same
degree of accuracy. For this task, power series expansions will
work nicely. 

In the following subsections, we will apply this procedure for the computation of the center manifold theory for certain critical points.

\subsection{Horizon ($P_H$) for massless scalar field}
\label{PHmassless}
We introduce the new variables 
\begin{subequations}
\begin{eqnarray}
& u= X_2+X_3-1, \\
& v= X_3-1,
\end{eqnarray}
\end{subequations} 
that translates the point $P_H$ to the origin. 
Then, using the center manifold theorem \cite{wiggins}, we find that the center manifold of the origin is given locally by the graph
\begin{eqnarray}
\Big\{(u, v)\in\mathbb{R}^2: v=\frac{1}{2} \left(u+\sqrt{1-u (u+2)}-1\right),~|u|\ll 1 \Big\}. 
\end{eqnarray}
In the original variables, the center manifold of $P_{H}$ is an arc of the circle $X_2^2+X_3^2=1$. 
Since it is an invariant set, the equation on the center manifold is $\tilde{u}=0$, as expected. 
The center manifold is unstable as shown in Fig. \ref{fig:Fig1}.

\subsection{Singularity ($P_S$) for massless scalar field}
\label{PSmassless}
The new variables that translates the point $P_S$ to the origin are
\begin{subequations}
\begin{eqnarray}
& u= \frac{2}{5} \left(3 X_2+X_3-\sqrt{5}\right), \\
& v=\frac{3}{5} \left(-2 X_2+X_3+\sqrt{5}\right).
\end{eqnarray}
\end{subequations} 
Hence, the center manifold of the origin is given locally by the graph
\begin{eqnarray}
\Big\{(u, v)\in\mathbb{R}^2: v=\frac{3 \left(2-\sqrt{-5 u^2-4 \sqrt{5} u+4}\right)}{4 \sqrt{5}}-\frac{3 u}{4},~|u|\ll 1\Big\}. 
\end{eqnarray}
In the original variables, the center manifold of $P_{S}$ is an arc of the circle $X_2^2+X_3^2=1$. 
Since it is an invariant set, the equation on the center manifold is $\tilde{u}=0$, as expected. The center manifold of $P_S$ is unstable as shown in Fig. \ref{fig:Fig1}.  

\subsection{Horizon ($P_H$) for exponential potential}
\label{PHexp}
As before, introducing the new variables that translates the point $P_H$ to the origin
\begin{subequations}
\begin{eqnarray}
& u=Y_{1}-\frac{\lambda  (X_{2}+X_{3}-1)}{\sqrt{2}},\\
& v_1=\frac{\lambda  (X_{2}+X_{3}-1)}{\sqrt{2}},\\
& v_2=X_{3}-1,
\end{eqnarray}
\end{subequations} 
we get the center manifold of the origin is given by the graph
\begin{eqnarray}
&\left\{(u, v_1, v_2)\in\mathbb{R}^3: v_1=h_1(u),~v_2=h_2(u), \right. \nonumber \\ & \left.
2 h_1(u) \left(\lambda  u-\sqrt{2} h_2(u)\right)+\left(\lambda +\frac{2}{\lambda }\right) h_1(u)^2 \right. \nonumber \\ & \left. +\lambda  \left(2 h_2(u)
   (h_2(u)+1)+u^2\right)=0, \right. \nonumber \\ & \left. \sqrt{1-\left(h_2(u)-\frac{\sqrt{2} h_1(u)}{\lambda }\right)^2}-\frac{\sqrt{2} h_1(u)}{\lambda }=1\right\}. 
\end{eqnarray}
This result have been confirmed by using Taylor series up to order $\mathcal{O}(|u|^{12})$ by substituting $h_1=-\frac{\lambda  u^4}{8 \sqrt{2}}-\frac{\lambda  u^6}{16 \sqrt{2}}+\frac{\lambda ^2 u^7}{32}-\frac{5 \lambda  u^8}{64 \sqrt{2}}+\frac{5 \lambda ^2 u^9}{128}-\frac{\left(\lambda 
   \left(11 \lambda ^2+42\right)\right) u^{10}}{512 \sqrt{2}}+\frac{9 \lambda ^2 u^{11}}{128}+\mathcal{O}\left(u^{12}\right)
	~\text{and}~h_2=-\frac{u^2}{2}-\frac{u^4}{4}+\frac{\lambda  u^5}{8 \sqrt{2}}-\frac{3
   u^6}{16}+\frac{3 \lambda  u^7}{16 \sqrt{2}}+\frac{1}{256} \left(-9 \lambda ^2-50\right) u^8+\frac{9 \lambda  u^9}{32 \sqrt{2}}+\frac{1}{128} \left(-11 \lambda ^2-28\right) u^{10}+\frac{\lambda  \left(13 \lambda ^2+218\right) u^{11}}{512 \sqrt{2}}+\mathcal{O}\left(u^{12}\right)$.

In the original variables, the center manifold of $P_{H}$ is a subspace of the invariant manifold $M:~X_2+X_3-\sqrt{1-X_2^2}=0,~X_2^2+X_3^2+Y_1^2=1$ represented by the red line in Fig. \ref{Fig:exp3D}. Since it is an invariant set, the equation on the center manifold is $\tilde{u}=0$, as expected. Besides, it is easy to conclude that we have only one trajectory, the line $X_2+X_3-\sqrt{1-X_2^2}=0, ~X_2^2+X_3^2=1,~0\leq X_2\leq \frac{1}{\sqrt{2}}$, and therefore one solution connecting the Horizon $P_H$ to Minkowski point $P_M$.

\subsection{Singularity ($P_S$) for exponential potential}
\label{PSexp}
Introducing the new variables that translates point $P_S$ to the origin
\begin{subequations}
\begin{eqnarray}
& u=\frac{\lambda  \left(-3 X_2-X_3+\sqrt{5}\right)}{3 \sqrt{2}}+Y_1,\\
& v_1=\frac{3}{5} \left(-2 X_2+X_3+\sqrt{5}\right),\\
& v_2=\frac{\lambda \left(3 X_2+X_3-\sqrt{5}\right)}{3 \sqrt{2}},
\end{eqnarray}
\end{subequations} 
we find that the center manifold of the origin is the graph
\begin{eqnarray}
&\left\{(u, v_1, v_2)\in\mathbb{R}^3: v_1=h_1(u),~v_2=h_2(u), \right. \nonumber \\ & \left. h_1(u) \left(\frac{2 \sqrt{2} h_2(u)}{\lambda }-\frac{2 \sqrt{5}}{3}\right)+\frac{10 h_1(u)^2}{9} \right. \nonumber \\ & \left. +h_2(u) \left(\frac{18 h_2(u)}{5 \lambda
   ^2}+h_2(u)+2 u\right)+u^2=0, \right. \nonumber \\ & \left. \frac{1}{15} \Big(-15 \sqrt{1-\frac{\left(-5 \lambda  h_1(u)+9 \sqrt{2} h_2(u)+6 \sqrt{5} \lambda \right)^2}{225 \lambda ^2}}  \right. \nonumber \\ & \left. +10
   h_1(u)+\frac{27 \sqrt{2} h_2(u)}{\lambda }+3 \sqrt{5}\Big)=0\right\}. 
\end{eqnarray}
This result have been confirmed by using Taylor series up to order $\mathcal{O}(|u|^{12})$ by substituting $h_1=\frac{3 u^2}{2 \sqrt{5}}+\frac{3 u^4}{4 \sqrt{5}}-\frac{\lambda  u^5}{8 \sqrt{2}}+\frac{9 u^6}{16 \sqrt{5}}-\frac{3 \lambda  u^7}{16 \sqrt{2}}+\frac{3}{256} \sqrt{5}
   \left(\lambda ^2+10\right) u^8-\frac{9 \lambda  u^9}{32 \sqrt{2}}+\frac{\left(55 \lambda ^2+252\right) u^{10}}{384 \sqrt{5}}-\frac{\left(\lambda  \left(65 \lambda
   ^2+1962\right)\right) u^{11}}{4608 \sqrt{2}}+\mathcal{O}\left(u^{12}\right) ~\text{and}~
	h_2=-\frac{1}{24} \left(\sqrt{\frac{5}{2}} \lambda \right) u^4-\frac{1}{48} \left(\sqrt{\frac{5}{2}} \lambda \right)
   u^6+\frac{5 \lambda ^2 u^7}{288}-\frac{5}{192} \left(\sqrt{\frac{5}{2}} \lambda \right) u^8+\frac{25 \lambda ^2 u^9}{1152}-\frac{\left(\sqrt{\frac{5}{2}} \lambda  \left(55 \lambda^2+378\right)\right) u^{10}}{13824}+\frac{5 \lambda ^2 u^{11}}{128}+\mathcal{O}\left(u^{12}\right)$.

In the original variables, the center manifold of $P_{S}$ is a subspace of the invariant manifold $M: X_2+X_3-\sqrt{1-X_2^2}=0,~X_2^2+X_3^2+Y_1^2=1$ represented by the red line in Fig. \ref{Fig:exp3D}. Since it is an invariant set, the equation on the center manifold is $\tilde{u}=0$, as expected. Besides, it is easy to conclude that we have only one trajectory, the line $X_2+X_3-\sqrt{1-X_2^2}=0, X_2^2+X_3^2=1, -\frac{1}{\sqrt{5}}<X_3< 0$, and therefore one solution connecting the Singularity $P_S$  to Minkowski point $P_M$. 

\subsection{Horizon ($P_H$) for arbitrary potential}
\label{centerPH}
To investigate the stability of the horizon $P_H$, we introduce the linear transformation 
\begin{subequations}
\begin{eqnarray}
\label{expansion:PH}
&u_1=\frac{1}{2} (2 (\lambda -\lambda_{c})+\lambda_{c} (X_2+X_3-1)),\\
&u_2=\lambda_{c} (X_{2}+X_{3}-1)-\sqrt{2}Y_{1},\\
&v_1=1-X_{2}-X_{3},\\
&v_2=1-X_{3},
\end{eqnarray}
\end{subequations}	
that translates a point on the line $P_H$ for a given $\lambda_c$ to the origin. 
We find that the center manifold of the origin can be expressed as
\begin{align}
\left\{(u_1,u_2,v_1,v_2)\in\mathbb{R}^4: v_1=\frac{u_2^4}{32}+\mathcal{O}(5), ~v_2=\frac{u_2^2}{4}+\frac{u_2^4}{16}+\mathcal{O}(5),~|u_2|<\delta\right\},~ \delta\ll 1. 
\end{align}
where $\mathcal{O}(5)$ denotes terms of fifth order.  
The center manifold of $P_H$ is the surface defined by $X_2+X_3-\sqrt{1-X_2^2}=0, X_2^2+X_3^2+Y_1^2=1, \lambda\in\mathbb{R}$, which is confirmed by substituting in these equations the following asymptotic expressions (found from \ref{expansion:PH})
\begin{eqnarray}
&X_2=\frac{u_2^2}{4}+\frac{u_2^4}{32}+\mathcal{O}(5),\\
&X_3=1-\frac{u_2^2}{4}-\frac{u_2^4}{16}+\mathcal{O}(5),\\
&Y_1=-\frac{u_2}{\sqrt{2}}-\frac{\lambda_c u_2^4}{32\sqrt{2}}+\mathcal{O}(5),\\
&\lambda=\lambda_c +u_1+\frac{\lambda_{c} u_2^4}{64}+\mathcal{O}(5).
\end{eqnarray}
The dynamics on the center manifold is given by 
\begin{subequations}
\begin{eqnarray}
\label{equ}
\tilde u_1=&u_2 f({\lambda_c})+u_1 u_2 f'({\lambda_c})+\frac{1}{2} u_1^2 u_2 f''({\lambda_c})+\frac{1}{6} u_1^3 u_2 f^{(3)}({\lambda_c}) \nonumber \\
&+\frac{1}{32}
   {\lambda_c} u_2^4 f({\lambda_c})+\mathcal{O}(5),\\
\tilde u_2= &\mathcal{O}(5).	
\end{eqnarray}
\end{subequations}
Now, let us find the non-linear terms which are relevant for the dynamics. 
Assuming that $f(\lambda_c)\neq 0$ and considering the change of coordinates $x=u_1, y= f(\lambda_c)u_2$ the above system corresponds to the Takens-Bogdanov normal form  with linear part 
$J=\left(\begin{array}{cc}
0 & 1\\
0& 0
\end{array}\right)$ \cite{Bogdanov,wiggins}.

Under the successive coordinate transformations
\begin{subequations}
\begin{eqnarray}
& (x,y)\rightarrow \left(x+\frac{x^2 f'(\lambda_{c})}{2 f(\lambda_{c})}, y\right)
\\
& (x,y)\rightarrow \left(x+\frac{x^3 \left(f(\lambda_c) f''(\lambda_c)+f'(\lambda_c)^2\right)}{6 f(\lambda_c)^2}+\frac{x^2 y}{2},  y+x y^2\right)
\end{eqnarray}
\end{subequations}
we obtain the normal form
\begin{subequations}
\label{A9}
\begin{eqnarray}
&\tilde x=y+\mathcal{O}(4),\\
&\tilde y=-y^3+\mathcal{O}(4).
\end{eqnarray}
\end{subequations}
where $\tilde f \equiv \frac{df}{d\xi}, \text{($\xi$ is an affine parameter)}$, and the higher terms are 
\newline
$\left(\frac{x^3 y \left(f(\lambda_c)^2 f^{(3)}(\lambda_c)-3 f'(\lambda_c)^3\right)}{6 f(\lambda_c)^3}+\frac{\lambda_c
   y^4}{32 f(\lambda_c)^3},0\right)+\mathcal{O}(5)$. It can be proven from these normal forms the instability of the horizon critical point $P_H$.
%
%
%
%
%
\subsection{Singularity ($P_S$) for arbitrary potential}
\label{centerPS}
To investigate the stability of the singularity $P_S$, we introduce the linear transformation 
\begin{subequations}
\begin{eqnarray}
&u_1=\lambda -\lambda_{c}+3 X_2+X_3-\sqrt{5},\\
&u_2=\lambda_{c} \left(3 X_2+X_3-\sqrt{5}\right)+Y_1,\\
&v_1=-2 X_{2}+X_{3}+\sqrt{5},\\
&v_2=-3 X_{2}-X_{3}+\sqrt{5},
\end{eqnarray}
\end{subequations}	
that translates a point on the line $P_S$ for a given $\lambda_c$ to the origin. 
The center manifold of the origin can be expressed as
\begin{align}
\left\{(u_1,u_2,v_1,v_2)\in\mathbb{R}^4: v_1=\frac{\sqrt{5} u_2^4}{4}+\frac{\sqrt{5} u_2^2}{2}+\mathcal{O}(5),~v_2=\frac{\sqrt{5} u_2^4}{8}+\mathcal{O}(5),~|u_2|<\delta\right\}, \quad \delta\ll 1, 
\end{align}
where $\mathcal{O}(5)$ denotes terms of fifth order on the vector norm. Hence, we have the following asymptotic expressions
\begin{eqnarray}
&X_2=\frac{2}{\sqrt{5}}-\frac{u_2^2}{2 \sqrt{5}}-\frac{3 u_2^4}{8 \sqrt{5}}+\mathcal{O}(5),\\
&X_3=-\frac{1}{\sqrt{5}}+\frac{3 u_2^2}{2
   \sqrt{5}}+\frac{u_2^4}{2 \sqrt{5}}+\mathcal{O}(5),\\
&Y_1=u_2+\frac{1}{8} \sqrt{5} \lambda_{c} u_2^4+\mathcal{O}(5),\\
&\lambda=\lambda_c +u_{1}+\frac{\sqrt{5}u_2^4}{8}+\mathcal{O}(5),
\end{eqnarray}
where $u_1$ is found by integrating \eqref{equ} and truncating the solution up to order $u_2^4$. 
The center manifold of $P_S$ is a subspace of $X_2+X_3-\sqrt{1-X_2^2}=0, X_2^2+X_3^2+Y_1^2=1, \lambda\in\mathbb{R}$.

The dynamics on the center manifold is given by 
\begin{subequations}
\begin{eqnarray}
\label{equ2}
\tilde u_1=&-\sqrt{2} u_2 f(\lambda_c)-\sqrt{2} u_1 u_2
   f'(\lambda_c)-\frac{u_1^2 u_2 f''(\lambda_c)}{\sqrt{2}}-\frac{u_1^3 u_2 f^{(3)}(\lambda_c)}{3 \sqrt{2}} \nonumber \\
	&-\frac{1}{4} \sqrt{\frac{5}{2}} \lambda_c u_2^4 f(\lambda_c)+\mathcal{O}(5),\\
\tilde u_2=& \mathcal{O}(5).	
\end{eqnarray}
\end{subequations}
For $f(\lambda_c)\neq 0$ and under the change of coordinates $x=u_1,y=-\sqrt{2} u_2 f(\lambda_{c})$ we have, as before, the Takens-Bogdanov normal form  with linear part 
$J=\left(\begin{array}{cc}
0 & 1\\
0& 0
\end{array}\right)$ \cite{Bogdanov,wiggins}. The analysis proceeds exactly as in \ref{centerPS} to obtain the normal form \eqref{A9}. 

\subsection{Anti-de Sitter point for arbitrary potentials}
\label{PAdS}
We know from our previous analysis that the Anti-de Sitter point ($P_{AdS}$) exists for the choice $V (\psi) < 0$ and it is stable for $f(0) > 0$. But this condition is only a sufficient condition, it is not a necessary condition. Now we will clarify what happens at the bifurcation value $f(0)=0$.\\
Under the linear transformation 
\begin{subequations}
\begin{eqnarray}
&u= \lambda,\\
&v_1= Y_1-\frac{\lambda }{2},\\
&v_2= -\frac{3 X_2}{2}-\frac{X_3}{2}+\sqrt{2},\\
&v_3=\frac{3}{2}
   \left(X_2+X_3-\sqrt{2}\right),
	\end{eqnarray}
	\end{subequations}
and using the center manifold theorem \cite{wiggins}  we find that the center manifold of the origin is given approximately by the graph
\begin{eqnarray}
&\left\{(u, v_1,v_2,v_3)\in\mathbb{R}^4: v_1= a_1 u^2+ a_2 u^3+a_3
   u^4+\mathcal{O}\left(u^5\right), \right. \nonumber \\ & \left. v_2= b_1 u^2+ b_2 u^3+b_3
   u^4+\mathcal{O}\left(u^5\right), \right. \nonumber \\ & \left. v_3= c_1 u^2+ c_2 u^3+c_3
   u^4+\mathcal{O}\left(u^5\right)\right\}. 
\end{eqnarray}
where $a_1= \frac{f'(0)}{6},\; a_2=\frac{1}{48} \left(4 f''(0)+8 f'(0)^2-3\right),$ $a_3=\frac{12 f^{(3)}(0)+112 f'(0)^3+3 f'(0) \left(28
   f''(0)-5\right)}{432},$ $b_1=\frac{1}{4 \sqrt{2}},\; b_2=\frac{f'(0)}{6 \sqrt{2}},\; b_3=\frac{48 f''(0)+112 f'(0)^2-27}{576 \sqrt{2}},\; c_1=-\frac{3}{8
   \sqrt{2}},\; c_2=-\frac{f'(0)}{4 \sqrt{2}},\; c_3=-\frac{48 f''(0)+112 f'(0)^2-27}{384 \sqrt{2}}$. 
In the original variables, the center manifold of $P_{AdS}$ is 
\begin{eqnarray}
\left\{(X_2,X_3,Y_1)\in\mathbb{R}^3: X_{3}-\frac{\sqrt{1-Y_1^2}}{\sqrt{2}}=0,~X_2-X_3=0\right\}.
\end{eqnarray}
The evolution on the center manifold is governed by the approximated equation 
\begin{eqnarray}
\label{B.3}
&u'=-\frac{u^2 f'(0)}{\sqrt{2}}-\frac{u^3 \left(3 f''(0)+2 f'(0)^2\right)}{6 \sqrt{2}}\nonumber \\
&+\frac{u^4 \left(-4 f^{(3)}(0)-8 f'(0)^3+f'(0) \left(3-8 f''(0)\right)\right)}{24
   \sqrt{2}}+\mathcal{O}\left(u^5\right)
\end{eqnarray}
For any potential satisfying $f'(0)=\pm \frac{3}{2\sqrt{2}}, f''(0)=-\frac{3}{4}, f^{(3)}=0$, 
the evolution equation \eqref{B.3} simplifies to 
\begin{eqnarray}
&u'=\mp\frac{3 u^2}{4}+O\left(u^5\right).
\end{eqnarray}
The equilibrium point $u= 0$, satisfy $f'(0) = 0$. For the positive sign in the above equation, the solutions with $u(0) < 0$ approach the equilibrium as time passes, while solutions with $u(0) > 0$ leave it (and go off to $\infty$ in finite time). Concerning the equation with negative sign, if we take the time reversal $t \rightarrow -t$, we end up with the same case and have the same results upon a time reversal. Such an equilibrium with one-sided stability is sometimes said to be semi-stable. (Example 2.3 in \cite{Davis}).
Now, for the choice $f(\lambda)=-\frac{1}{N} \lambda^2$ (power-law potential $\psi^{N}$), the evolution equation \eqref{B.3} simplifies to 
\begin{eqnarray}
&u'=\frac{u^3}{\sqrt{2}N}+O\left(u^5\right).
\end{eqnarray}
The equilibrium $u = 0$ is asymptotically stable for $N<0$, while for $N>0$ it is unstable, even though $f'(0) = 0$ in both cases. Furthermore, perturbations from the equilibrium grow or decay algebraically in time, not exponentially as in the linear stability analysis (Example 2.4 in \cite{Davis}).
Finally, in the general case, and if we neglect the fifth order terms, we obtain a 1-dimensional gradient differential equation
\begin{eqnarray}
&u'=-\frac{d U(u)}{du}, \\
&U(u)=\frac{u^3 f'(0)}{3 \sqrt{2}}+\frac{u^4 \left(3 f''(0)+2 f'(0)^2\right)}{24 \sqrt{2}} \nonumber \\
&+\frac{u^5 \left(4 f^{(3)}(0)+8 f'(0)^3+f'(0) \left(8 f''(0)-3\right)\right)}{120 \sqrt{2}},
\end{eqnarray} where $U$ is the effective potential.  
Depending on whether $u=0$ is (i) an inflection point of $U$, (ii) a (possible degenerated) minimum of $U$, or (iii) a (possible degenerated)  maximum of $U$, we will have (i) semi-stability (i.e.,  some orbits approach the equilibrium as time passes,
while other solutions departs from it, or blows up in finite time), (ii) stability or (iii) instability for the fixed point $u=0$. The previous examples captures the main features of this classification.  
\section{Flare-out condition}
\label{appendix:flare}

\begin{figure}[htb!]
\centering
\includegraphics[width=0.45\textwidth]{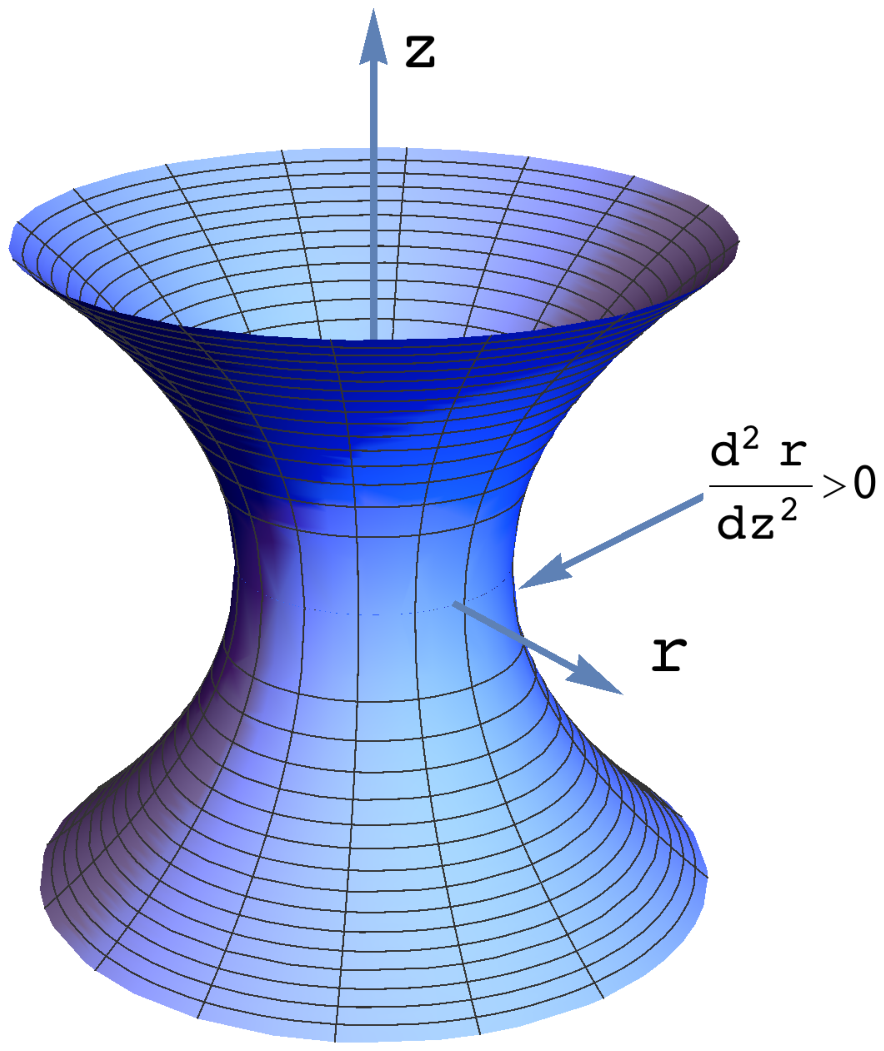}
\caption{\label{throat} Embedding of a wormhole and its throat in 3D euclidean space.}
\end{figure}

In this Appendix we examine the requirements for the satisfaction of the flare-out 
condition. Considering a metric of the following form
\begin{eqnarray}
{ds}^2=-A(r)dt^2+\frac{dr^2}{B(r)}+r^2d\theta^2+r^2\sin^2(\theta)d\phi^2,
\end{eqnarray}
which at $t=$const and $\theta=\pi/2$ reduces to 
\begin{eqnarray}
{ds}^2=\frac{dr^2}{B(r)}+r^2d\phi^2,
\end{eqnarray}
we can elaborate the flare-out condition by considering the embedding geometry. In fact 
this spacetime can be embedded in a 3D flat space with cylindrical coordinates
$(r,\phi,z)$, namely
\begin{eqnarray}
{ds}^2=\frac{dr^2}{B(r)}+r^2d\phi^2 &=dr^2+dz^2+r^2d\phi^2\nonumber\\
&=\Bigl[1+\Bigl(\frac{dz}{dr}\Bigr)^2\Bigr]dr^2+r^2 d\phi^2,
\end{eqnarray}
where
\begin{eqnarray}
\frac{dz}{dr}=\pm \sqrt{\frac{1-B(r)}{B(r)}}.
\end{eqnarray}
The space which is located at $B=0$ is a throat (minimum radius) if the flare-out 
condition is 
satisfied, namely if
\begin{eqnarray}
\frac{d^2r}{dz^2}>0.
\end{eqnarray}
However, since 
\begin{eqnarray}
\frac{d^2r}{dz^2}=\frac{x_2'}{r (1-x_2^2)^2},
\end{eqnarray}
we conclude that the flare-out condition is satisfied if and only if
\begin{eqnarray}
x_2'>0 \quad \text{when}~~~x_2=0.
\end{eqnarray}
In Fig. \ref{throat} we present the throat of the wormhole.

\section{Misner-Sharp mass}
\label{appendix:mass}

In this Appendix we examine the Misner-Sharp mass. Using the $1+1+2$ decomposition for 
LRS-II spacetimes, it can be shown that the Misner-Sharp mass 
takes the following form \cite{Ellis:2014jja}:
\begin{eqnarray}
\mathcal{M}=\frac{1}{2K^{3/2}}\Bigl(\frac{\rho}{3}-\mathcal{E}-\frac{\Pi}{2}\Bigr),
\end{eqnarray}
which can be written in our set of variables as
\begin{eqnarray}
\mathcal{M}&=\frac{1-x_2^2}{2\sqrt{K}}.
\end{eqnarray}
We see therefore that the Minkowski spacetime for which $x_2=1$ correspond to $\mathcal{M}=0$.\\

\end{document}